\newif\ifvldb
\vldbfalse

\ifvldb
\documentclass[dvipdfmx]{vldb}
\else
\documentclass{vldb_arxiv}
\fi

\usepackage{balance} 
\usepackage{caption}
\usepackage{subcaption}
\usepackage{graphicx}
\usepackage{ascmac}
\usepackage{multirow}
\usepackage{seqsplit}
\usepackage{color}
\usepackage{url}
\usepackage{cite}

\ifvldb
\vldbTitle{An Analysis of Concurrency Control Protocols for In-Memory Databases with CCBench}
\vldbAuthors{Takayuki Tanabe, Takashi Hoshino, Hideyuki Kawashima, Osamu Tatebe}
\vldbDOI{https://doi.org/10.14778/3424573.3424575}
\vldbVolume{13}
\vldbNumber{13}
\vldbYear{2020}
\else
\fi

\begin{document}
\ifvldb
\else
\title{An Analysis of Concurrency Control Protocols\\ for In-Memory Databases with CCBench\\(Extended Version)}
\fi
\ifvldb
\else
\author{
\alignauthor
Takayuki Tanabe$^1$\hspace{-0.08cm}\titlenote{Work done while at University of Tsukuba}\hspace{-0.2cm},
Takashi Hoshino$^2$,
Hideyuki Kawashima$^4$,\\
Jun Nemoto$^4$,
Masahiro Tanaka$^4$,
Osamu Tatebe$^3$\\
\affaddr{
$^1$NAUTILUS Technologies, Inc.,
$^2$Cybozu Labs, Inc.,
$^3$University of Tsukuba,
$^4$Keio University}\\
\affaddr{
$^1$tanabe@nautilus-technologies.com,
$^2$hoshino@labs.cybozu.co.jp,
$^3$tatebe@cs.tsukuba.ac.jp,\\
$^4$\{hideyuki.kawashima, nemoto, masa16.tanaka\}@keio.jp}
}
\fi

\pagestyle{empty}

\maketitle

\begin{abstract}
 This paper presents yet another concurrency control analysis platform, CCBench.
 CCBench supports seven protocols (Silo, TicToc, MOCC, Cicada, SI, SI with latch-free SSN, 2PL) and seven versatile optimization methods and enables the configuration of seven workload parameters.
 %
 We analyzed the protocols and optimization methods using various workload parameters and a thread count of 224.
 Previous studies focused on thread scalability and did not explore the space analyzed here.
 We classified the optimization methods on the basis of three performance factors: CPU cache, delay on conflict, and version lifetime.
 Analyses using CCBench and 224 threads, produced six insights.
 (I1) The performance of optimistic concurrency control protocol for a read-only workload rapidly degrades as cardinality increases even without L3 cache misses.
 (I2) Silo can outperform TicToc for some write-intensive workloads by using invisible reads optimization.
 (I3) The effectiveness of two approaches to coping with conflict (wait and no-wait) depends on the situation.
 (I4) OCC reads the same record two or more times if a concurrent transaction interruption occurs, which can improve performance.
 (I5) Mixing different implementations is inappropriate for deep analysis.
 (I6) Even a state-of-the-art garbage collection method cannot improve the performance of multi-version protocols if there is a single long transaction mixed into the workload.
 On the basis of I4, we defined the read phase extension optimization in which an artificial delay is added to the read phase. On the basis of I6, we defined the aggressive garbage collection optimization in which even visible versions are collected.
 The code for CCBench and all the data in this paper are available online at GitHub.
\end{abstract}

\ifvldb
\else
\\
\\
\fi
\section{Introduction}
\ifvldb
\else
\vspace{-0.5cm}
\fi
\subsection{Motivation}
Transacting processing systems containing thousands of CPU cores in a single server
have been emulated~\cite{yu2014staring}, implemented~\cite{kimura2015foedus, wang2016mostly},
and analyzed~\cite{10.1145/3399666.3399910}.
Along with these systems, a variety of concurrency control protocols for a single server~
\cite{
  kim2016ermia,   
  lim2017cicada,  
  tu2013speedy,   
  wang2016mostly, 
  yu2016tictoc}    
have been proposed in recent years for use in many-core architectures.
These modern protocols use a variety of optimization methods, and their behaviors depend on the workload characteristics (e.g., thread count, skew, read ratio, cardinality, payload size, transaction size, and read-modify-write or not).
Recent new proposal studies have compared these protocols with conventional ones~
\cite{
  kim2016ermia,   
  lim2017cicada,  
  tu2013speedy,   
  wang2016mostly, 
  yu2016tictoc,   
  Repair,
  Healing,
  cavalia_paper,
  ACC_Tang,
  kimura2015foedus,
  IC3,
  STOv2,
  Strife,
  BCC,
  EWV,
  AOCC_Guo,
  tx_batch,
  SGT}, and recent analytical studies have compared the performance of conventional protocols on a common platform~
\cite{
  trireme,                 
  10.1145/3399666.3399910, 
  wu2017empirical,         
  yu2014staring}.          
These studies mostly evaluated protocol scalability.
This paper acknowledges that modern protocols are scalable
and focuses on other factors that contribute to performance on a many-core environment.
This type of analysis is novel to the best of our knowledge.
%

Fairness is important in comparison. However,
some recent studies were unable to perform such analysis fairly because
they compared their new protocols with others using different platforms.
%
For example, evaluations of \mbox{ERMIA}~\cite{kim2016ermia}, mostly-optimistic concurrency control
(MOCC)\cite{wang2016mostly}, and Cicada~\cite{lim2017cicada} used two or three platforms.
Experiments using a mix of platforms can produce only approximate results because the performance of protocols on a many-core architecture depends greatly on the exploitation of the underlying hardware.
%
In such conditions,
only the macroscopic evaluations (i.e., scalability) can be conducted,
and a detailed analysis is difficult.
A single unified comparison platform is needed to conduct a detailed analysis.

%
%
For a fair comparison, a common analysis platform is necessary.
It should provide shared core modules such as access methods (concurrent index) and thread-local data structures (read and write sets) for any protocols.
It should also provide a variety of optimization methods
to enable obtaining a deep understanding of the protocols.
Finally, the platform should be publicly available for reproducing the experiments.
Although there are several open-source platforms,
including DBx1000~\cite{dbx1000}, Peloton~\cite{peloton}, and Cavalia~\cite{cavalia},
they do not provide certain modern protocols~\cite{
  wang2017efficiently,   
  lim2017cicada,  
  wang2016mostly, 
  yu2016tictoc}.  
\ifvldb
\else
Appuswamy et al.~\cite{trireme} evaluated protocols in four types of architecture using Trireme, which is not publicly available.
  \fi
\subsection{Contributions}
\label{sec:Contribution}
The first contribution of this paper is {\bf CCBench}: a platform for fairly evaluating concurrency control protocols.
%
The protocols are two-phase locking (2PL), Silo, MOCC, TicToc, snapshot isolation (SI), latch-free serial safety net (SSN), and Cicada.
CCBench supports seven versatile optimization methods and two identified in this work (Table \ref{table::protocol_opt}, $\S$\ref{sec:A Platform for Fair Comparison}).
Optimization methods can thus be applied to non-original protocols.
For example, the {\sf NoWait} method~\cite{bernstein1987concurrency} can be applied to Silo, the rapid garbage collection (GC) method ~\cite{lim2017cicada} can be applied to multi-version protocols, and the {\sf AdaptiveBackoff} optimization method~\cite{lim2017cicada} can be applied to all protocols.
Fairness in analyzing protocols is achieved through shared modules including a workload generator, access methods (Masstree~\cite{masstree_beta, mao2012cache}), local datasets (read/write sets), and a memory manager (mimalloc~\cite{mimalloc} and {\sf AssertiveVersionReuse} presented in this paper).
Evaluation of protocols under various conditions is enabled by providing of seven workload configuration parameters: skew, payload size, transaction size, cardinality, read ratio, read-modify-write (RMW) or not, and the number of worker threads.
CCBench and all of the data in this paper are available online at GitHub~\cite{thawkccbench}.

The second contribution is {\bf an analysis of cache, delay, and version lifetime using 224 threads.}
We clarified the effects of the protocols on the performance factors by configuring the optimization methods and workload settings.
As suggested elsewhere~\cite{yu2014staring} and \cite{kimura2015foedus}, the era of a thousand cores is just around the corner, so conventional analytical studies focused on evaluating thread scalability~
\cite{
  10.14778/3055540.3055548,
  wu2017empirical,
  yu2014staring}.
In contrast, we performed all analyses with {\it 224 threads on 224 cores}.

We first investigated the effects of the optimization methods related to {\bf cache}.
By determining that a centralized counter increases cache-line conflicts and
degrades performance, we gained two insights.
{\bf I1:} The performance of optimistic concurrency control (OCC) for a read-only workload rapidly degrades as cardinality increases even without L3 cache misses ($\S$\ref{sec:Cache Miss Mitigation}).
{\bf I2:} Silo outperforms TicToc for write-intensive workloads by using {\sf InvisibleReads} ($\S$\ref{sec:Effect of Invisible Read}).

%
%
We then investigated the effects of the optimization methods related to {\bf delay},
and gained two more insights.
{\bf I3:} The effectiveness of two approaches to cope with conflict ({\sf Wait} and {\sf NoWait}) depends on the situation ($\S$\ref{sec:Confirmation of Effect of NoWait}).
%
{\bf I4:} OCC reads the same record two or more times due to concurrent transaction interruption.
Surprisingly, OCC can improve performance in certain situations with it, and we have defined a new optimization method {\sf ReadPhaseExtension} based on it ($\S$\ref{sec:Effect of Extra Read and Read Phase Extension Method}).

%
%
Finally, we investigated the effects of optimization methods related to {\bf version lifetime}
 for multi-version concurrency control (MVCC) protocols, and gained two final insights.
{\bf I5:} Mixing different implementations is inappropriate.
Silo outperforms Cicada on the Yahoo! Cloud Serving Benchmark B (YCSB-B) workload in an unskewed environment,
which is inconsistent with previously reported testing results on different systems~\cite{lim2017cicada}.
{\bf I6:} Even a state-of-the-art GC technique cannot improve the performance of MVCC if there is a single long transaction mixed into the workload.
To overcome this problem, we defined a new optimization method, {\sf AggressiveGC}.
It requires an unprecedented protocol that weaves GC into MVCC, thus going beyond the current assumption that
versions can not be collected if they might be read by transactions ($\S$\ref{sec:discussion-towards-version-lifetime-control}).

\subsection{Organization}
The rest of this paper is organized as follows.
$\S$\ref{sec:Preparation} reviews existing protocols.
$\S$\ref{sec:Platform} discusses the fairness condition and presents CCBench.
$\S$\ref{sec:Reproduction of Prior Work} describes the reproduction of experiments.
$\S$\ref{sec:Cache Effect} investigates the effect of cache.
$\S$\ref{sec:Delay Control} investigates the effect of delay.
$\S$\ref{sec:version-lifetime-control} investigates the effect of version lifetime management.
$\S$\ref{sec:Related Work} reviews related work, and
$\S$\ref{sec:Conclusion} concludes this paper and shows future directions.

\section{Preliminaries}
\label{sec:Preparation}
\subsection{Concurrency Control Protocols}
\label{sec:Concurrency Control Protocols}

The concurrency control protocols we analyzed are classified as
(1) pessimistic (2PL~\cite{eswaran1976notions}),
(2) optimistic (Silo~\cite{tu2013speedy} and TicToc~\cite{yu2016tictoc}),
(3) multi-version (SI~\cite{jung2013performance} and ERMIA~\cite{kim2016ermia}),
(4) integration of optimistic and pessimistic (MOCC~\cite{wang2016mostly}),
and (5) integration of optimistic and multi-version (Cicada~\cite{lim2017cicada}).

{\bf Silo~\cite{tu2013speedy}} is an OCC protocol that has influenced subsequent concurrency control protocols.
For example, FOEDUS~\cite{kimura2015foedus} and MOCC~\cite{wang2016mostly} extend the commit protocol of Silo.
Silo selects the design of {\sf InvisibleReads}~\cite{marathe2004design} so that it does not update the metadata during a read operation.
The invisible read process avoids cache-line conflicts, so it provides scalability, as shown by Wang and Kimura~\cite{wang2016mostly}.

{\bf TicToc~\cite{yu2016tictoc}}\label{protocol::tictoc} is an OCC protocol based on timestamp ordering with data-driven timestamp management.
TicToc has a larger scheduling space than Silo, so it can commit schedules where Silo cannot.
TicToc provides three optimization methods.
{\sf PreemptiveAbort} initiates abort processing immediately if an abort is detected before write locking in the validation phase.
{\sf NoWaitTT} does not wait for lock release in the validation phase.
It instead releases the locks and retries the validation phase after a fixed duration of a wait without an abort.
{\sf TimestampHistory} expands the scheduling space by recording the write timestamp of an older version so that some of the read operations on the version can be verified after the record is overwritten.

{\bf MOCC~\cite{wang2016mostly}} exhibits high performance for a wide range of workloads by adaptively switching its policy by adding pessimistic locking and temperature management to Silo.
MOCC locks high temperature (on highly conflicted) records while keeping the order of records locked to avoid deadlocks in the read phase.
The MOCC queuing lock (MQL) integrates an MCS (Mellor-Crummey and Scott) lock, which can time out~\cite{scott2001scalable}, and a reader/writer fairness MCS lock~\cite{mellor1991scalable}.

{\bf Cicada~\cite{lim2017cicada}} combines the OCC protocol, 
a multi-version protocol with timestamp ordering, and distributed timestamp generation.
The distributed timestamp generation eliminates the need to access the centralized counter that conventional MVCC protocols require, thereby dramatically mitigating cache-line conflicts.
Cicada has a variety of optimization methods.
{\sf EarlyAbort} initiates an abort if one is predicted during the read phase.
{\sf BestEffortInlining} embeds an inline version in the record header at the top of the version list to reduce indirect reference cost.
{\sf PrecheckValidation} checks whether read versions are still valid in the early validation phase.
{\sf RapidGC} is a quick and parallel GC optimization method.
{\sf AdaptiveBackoff} dynamically determines how long to wait before retrying.
{\sf SortWriteSetByContention} detects conflicts at an early stage before performing actions unnecessarily due to aborts.

{\bf SI~\cite{berenson1995critique}}
is an MVCC protocol that can generate write-skew and read-only-transaction anomalies, so it does not produce serializable schedules.
Under SI, a transaction reads a snapshot of the latest committed version.
The write operations are also reflected in the snapshot.
%
SI requires a monotonic-increasing timestamp assignment for each transaction to provide snapshots.
To determine the timestamp, a centralized shared counter is required.

{\bf SI with latch-free SSN~\cite{kim2016ermia}}~
integrates SI and SSN~\cite{wang2015serial} and exploits the latch-free mechanism~\cite{wang2017efficiently}.
SSN detects and aborts dangerous transactions that could lead to non-serializable schedules.
We refer to the integration of SI and SSN as {\bf ERMIA} in this paper.

{\bf 2PL~\cite{eswaran1976notions,  weikum2001transactional}}~
releases all read/write locks at the end of each transaction.
The {\sf NoWait} method, which immediately aborts the running transaction when a conflict is detected,
was originally proposed as a deadlock resolution mechanism~\cite{10.1145/356842.356846}.
We use it as an optimization method.

\subsection{Environment}
\label{sec:Environment}
The evaluation environment consisted of a single server with four Intel (R) Xeon (R) Platinum 8176 CPUs with 2.10GHz processors.
Each CPU had 28 physical cores with hyper-threading, and the server had 224 logical cores.
Each physical core had 32 KB private L1d cache and 1 MB private L2 cache.
The 28 cores in each processor shared a 38.5 MB L3 cache.
The total cache size was about 154 MB.
Forty-eight DDR4-2666 32 GB DIMMs were connected, so the total size was 1.5 TB.

In all graphs in this paper showing the results of CCBench, each plot point shows the average for five runs, each with more than 3 s.
We confirmed that these numbers produce stable CCBench results.
The error bars indicate the range between the maximum and minimum values.
We counted the number of committed and aborted transactions to calculate average throughput.
We used \texttt{rdtscp} instruction to measure the latency of the transitions and their portions.
Each worker thread was pinned to a specific CPU core in every experiment.

\section{CCBench}
\label{sec:Platform}

\begin{figure}[tb]
  \begin{center}
    \includegraphics[scale=0.46]{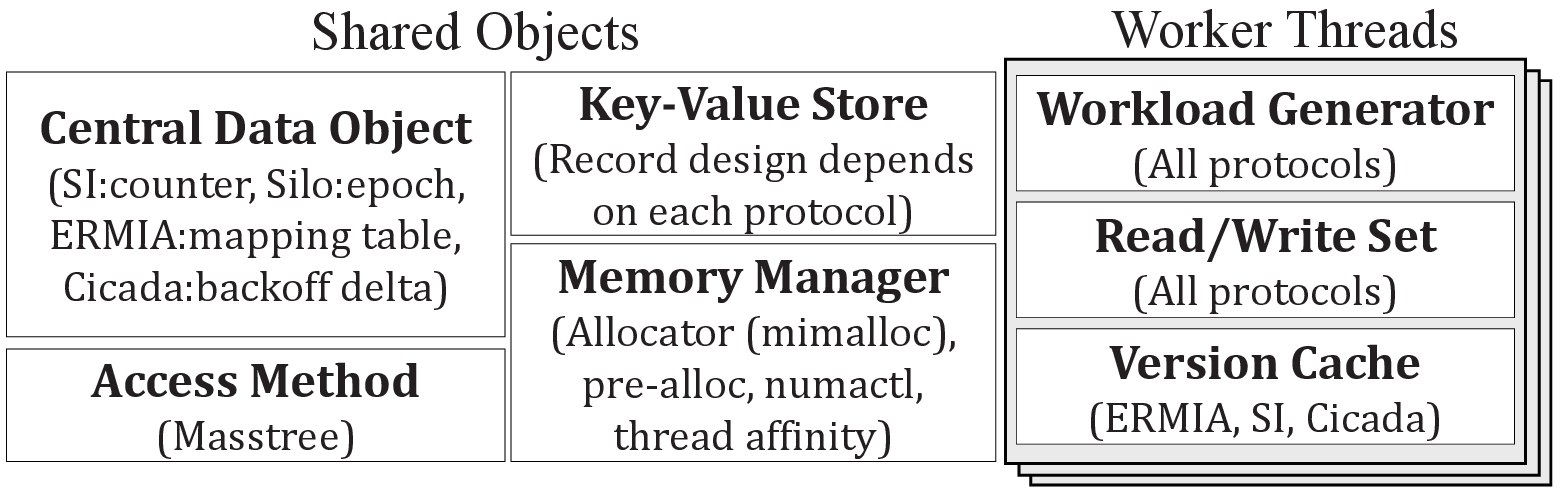}
    \caption{CCBench Architecture.}
    \label{figure::CCBench Architecture}
    \end{center}
\end{figure}

\subsection{Fairness Condition}
When analyzing concurrency control protocols, their performance should be evaluated under
a fairness condition.
The meaning of fairness depends on the situation.
In this paper, fairness means that {\bf basic modules are shared in the analysis}.
This is because the performance of modern protocols is closely related to the exploitation of the underlying hardware; i.e., they are sensitive to the engineering details.
Therefore, the basic modules of an evaluation platform should be shared for a fair evaluation.
Access methods (e.g., Masstree) and local data structures (read and write sets) need to be shared among protocols.
The effects of memory allocation should be reduced as much as possible.
The workload should be consistent among experiments.

%
%
Several analysis studies satisfy our fairness condition.
Yu et al.~\cite{yu2014staring} developed DBx1000~\cite{dbx1000}.
It was used in the evaluation of TicToc paper~\cite{yu2016tictoc} and in another evaluation with a real 1568 cores machines~\cite{10.1145/3399666.3399910}.
%
DBx1000 currently does not support some modern protocols (Cicada, MOCC, ERMIA).
Wu et al. developed Cavalia~\cite{cavalia} to compare the
hardware transactional memory (HTM)-assisted OCC-style protocol (HTCC)~\cite{cavalia_paper} with conventional protocols. Several modern protocols (Cicada, ERMIA, TicToc, MOCC) were beyond their scope.
Wu et al.~\cite{wu2017empirical} developed Peloton~\cite{peloton} to compare MVCC protocols.
Several modern protocols (Cicada, ERMIA, Silo, TicToc, MOCC) were beyond their scope.
Appuswamy et al.~\cite{trireme} developed Trireme to compare protocols in four types of architecture, including shared-everything.
Several modern protocols (TicToc, Cicada, MOCC, ERMIA) were beyond their scope.

A protocol can include a variety of optimization methods.
Even if protocol {\sf P} does not initially include optimization method {\sf O},
an analysis platform should provide {\sf O} to {\sf P} if users request it
because the performance of a protocol greatly depends on the optimization method.
For example, {\sf RapidGC} included in Cicada can also be applied to both SI and ERMIA.
{\sf NoWait}~\cite{10.1145/356842.356846} can be applied to Silo to improve performance,
as shown in Fig. \ref{fig:ycsbA_tuple100m_skew09_slprp0-1000_tps}.
Conventional platforms do not support this concept.

Our fairness condition was not satisfied in several studies.
The Cicada paper ($\S$4.2)~\cite{lim2017cicada} states,
``Cicada and existing schemes in DBx1000 share the benchmark code,
but have separate data storage and transaction processing engines.''
The MOCC paper~\cite{wang2016mostly} states that the pure OCC, MOCC, and PCC (2PL variants) implementations used the FOEDUS system, the Dreadlock/WaitDie/BlindDie (2PL variants) implementations used the Orthrus system, and the ERMIA implementation used the ERMIA system, which means three benchmark systems were mixed.
The ERMIA paper~\cite{kim2016ermia} states that the Silo implementation used the Silo system while the SI/SSI implementation used the ERMIA system, which means two benchmark systems were mixed.
In studies using a mix of systems,
experts must conduct a huge amount of effort to conduct a fair comparison, which would not be easy, especially in a deep understanding of protocols or optimizations.

%
%
%
%
%
\subsection{CCBench: A Platform for Fair Analysis}
\label{sec:A Platform for Fair Comparison}
\begin{table*}
 \begin{center}
  \begin{small}
   \caption{Versatile optimization methods in CCBench.
   {\sf ---}: Irrelevant or Incompatible.
   {\sf Org}: Supported by original protocol.
   {\sf CCB}: Supported by CCBench.
   ({$\alpha$}) Delay inspired by extra reads in OCC ($\S$\ref{sec:Effect of Extra Read and Read Phase Extension Method}). Proposed in this work.
   ({$\beta$}) CCBench performs read lock for hot records and invisible reads for non-hot records.
   ({$\gamma$})  {\sf NoWait} locking optimization detects a write lock conflict and immediately aborts and retries the transaction. {\sf NoWaitTT} releases all locks and re-locks them.
   ({$\delta$}) Lightweight memory management. Proposed in this work ($\S$\ref{sec:Implementing Optimization Techniques}).
   ({$\epsilon$}) {\sf PreempiveAbort} and {\sf TimestampHistory} cannot be applied to others.
   ({$\zeta$})   {\sf SortWriteSetByContention}, {\sf PrecheckVersionConsistency}, {\sf EarlyAborts}, {\sf BestEfforInlining} cannot be applied to others.
   We applied optimization methods as follows.
   {\sf NoWait}: Silo in Fig. \ref{fig:ycsbA_tuple100m_skew09_slprp0-1000_tps}.
   {\sf RapidGC}: ERMIA and SI in all cases.
   {\sf AssertiveVersionReuse}: Cicada, ERMIA, and SI in all cases. 
   {\sf AdaptiveBackoff}: TicToc in Fig. \ref{fig:tuple10m_val1k_ycsb_high_tps} and all protocols in Figs. \ref{fig:ycsbA_tuple10m_ope16_rmw_skew099_tps}, \ref{fig:ycsbA_tuple10m_ope16_rmw_skew0-099_th28_tps}, \ref{fig:ycsbB_tuple10m_ope16_rmw_skew0-099_th28_tps}, and \ref{fig:ycsbB_tuple10m_ope1_rmw_skew099_tps}.
   }
   \label{table::protocol_opt}
   \begin{tabular}[tb]{r|cc|ccc|cc}
    \begin{tabular}{c} {\rm Performance}\\{\rm Factor}\end{tabular}&
    \multicolumn{2}{|c|}{{\rm Cache}}&\multicolumn{3}{|c|}{{\rm Delay}}&\multicolumn{2}{|c}{{\rm Version Lifetime}}\\ \hline\hline
    \begin{tabular}{c} {\sf Optimization} \\{\sf Method}          \end{tabular}&
    \begin{tabular}{c} {\sf Decentralized}\\{\sf Ordering}        \end{tabular} &
    \begin{tabular}{c} {\sf Invisible}    \\{\sf Reads}           \end{tabular} &
    \begin{tabular}{c} {\sf NoWait}     \\or {\sf Wait}        \end{tabular} &
    \begin{tabular}{c} {\sf Adaptive}     \\{\sf Backoff}         \end{tabular} &
    \begin{tabular}{c} {\sf ReadPhase}    \\{\sf Extension}({$\alpha$})\end{tabular} &
    \begin{tabular}{c} {\sf AssertiveVersion}\\{\sf Reuse}({$\delta$})\end{tabular} &
    \begin{tabular}{c} {\sf Rapid}\\{\sf GC} \end{tabular} \\  \hline\hline
2PL~\cite{weikum2001transactional} &Org, CCB &---                 &Org,CCB       &CCB      &--- & --- &---\\
Silo~\cite{tu2013speedy}           &Org, CCB &Org, CCB            &CCB           &CCB      &CCB & --- &---\\
MOCC~\cite{wang2016mostly}         &Org, CCB &Org, CCB ({$\beta$})&---           &CCB      &CCB & --- &---\\
TicToc({$\epsilon$})~\cite{yu2016tictoc}&Org, CCB &---            &CCB ({$\gamma$}) &CCB   &CCB & --- &---\\
SI~\cite{jung2013performance}      &---      &---                 &---           &CCB      &--- & CCB &CCB\\
ERMIA~\cite{kim2016ermia}          &---      &---                 &---           &CCB      &--- & CCB &CCB\\
Cicada({$\zeta$})~\cite{lim2017cicada}       &Org, CCB &---       &---           &Org, CCB &CCB & CCB &Org, CCB\\
   \end{tabular}
  \end{small}
 \end{center}
\end{table*}

Our CCBench analysis platform for in-memory CC protocols satisfies our fairness condition
because it shares the basic modules among protocols. The architecture of CCBench is illustrated in Fig.~\ref{figure::CCBench Architecture}.
The code for CCBench is available on GitHub~\cite{thawkccbench}.

In CCBench, each thread executes both the client and the server logic.
The client generates a transaction with read/write operations using the workload parameters at runtime.
The server provides APIs as C++ functions, such as read, write, commit, and abort.
The client calls the APIs to run transactions.
The server runs the transactions requested from the client inside a worker thread.
The client code is separated from the server code, and the both codes are compiled into a single executable binary.
The memory allocator, mimalloc~\cite{mimalloc}, allocates interleaved memory among CPU sockets
by using the Linux {\tt numactl} command.
Memory allocation is avoided as much as possible so that the penalty~\cite{clements2013radixvm} imposed by the Linux virtual memory system can be ignored and so that the execution performance of protocols is not degraded due to undesirable side-effects of memory allocation.
CCBench initializes the database for each experiment,
pre-allocates the memory for objects, and reuses the memory.
Allocated memory for local data structures (read/write sets)
and the generated operations for a transaction are reused for the next transaction.
The metadata and record data are carefully aligned to reduce false sharing.
A wrapper of Masstree~\cite{masstree_beta} was implemented, and all protocols used it as the access method.

CCBench supports seven versatile optimization methods, as shown in Table~\ref{table::protocol_opt}.
(1) {\sf DecentralizedOrdering}: prevents contended accesses to a single shared counter.
(2) {\sf InvisibleReads}: read operations that do not update memory and cache-line conflicts do not occur.
(3) {\sf NoWait} or {\sf Wait}: immediate abort upon detecting a conflict followed by retry, or waiting for lock release.
(4) {\sf ReadPhaseExtension}: an artificial delay added to the read phase, inspired by the extra read process that retries the read operation.
(5) {\sf AdaptiveBackoff}: an artificial delay before restarting an aborted transaction.
(6) {\sf AssertiveVersionReuse}: allocates thread-local space, denoted as version cache in Fig. \ref{figure::CCBench Architecture}, to retain versions so that memory manager access is not needed.
(7) {\sf RapidGC}: frequent updating of timestamp watermark for GC in MVCC protocols.
{\sf ReadPhaseExtension} (4) and {\sf AssertiveVersionReuse} (6) are presented in this paper.

CCBench enables the use of additional optimization methods for 2PL, Silo, TicToc, SI, ERMIA, and Cicada.
%
It does not support the read-only optimization methods introduced in Silo, FOEDUS, MOCC, and ERMIA  because they improve performance only in a specific case
(i.e., 99\% read-only transactions, 1\% write transactions, skew at 0.99),
and such a workload is not considered here.

%
%
Users of CCBench can easily add new protocols to the platform.
After replicating a directory that has an existing protocol, the user rewrites the functions corresponding to the transactions with begin/read/write/commit/abort operations in the transaction executor class.
The user can then reuse the basic modules in the platform: workload generator, memory management, and so on.
Users can easily attach or detach optimization methods provided by CCBench to their protocols simply by rewriting the preprocessor definition in the CMakeLists.txt.
A developer guide for CCBench users is available~\cite{cc_format}.
One of our team members implemented a simple OCC protocol following this guide~\cite{occ_nemoto}; and published a description of the experience for other users~\cite{nemotoccbench}.
Users can also switch the configuration of the optimization methods, key-value size, and protocol details in the same way as done in the online example.
In an experiment, the user can set the workload by specifying the runtime arguments by gflags~\cite{gflags} using the seven workload parameters listed in Table \ref{tab:Experimental Parameters}.
%
%
This design makes it easier to conduct experiments than DBx1000~\cite{dbx1000}, which manages these values as preprocessor definitions in a file.


\subsection{Optimization Method Implementations}
\label{sec:Implementing Optimization Techniques}
{\bf 2PL:~}
For efficiency, we implemented from scratch the reader/writer lock system with the compare-and-swap (CAS) operation used in the protocol.

{\bf Silo:~} We padded the global epoch (64 bits) into the cache-line size to reduce the false sharing that can be caused by a global epoch state change.
Calculation of the commit transaction id (TID) is expensive since it requires loops for both the read set and the write set.
We reduced these two loops by implementing calculation at write lock acquisition and read verification.

{\bf ERMIA:~}
We used latch-free SSN to prevent expensive SSN serial validation.
This implementation avoids the validation from being a critical section.
We integrated pstamp and sstamp to reduce memory usage as described at Section 5.1 of original paper~\cite{wang2017efficiently}.
We optimized the transaction mapping table.
Straightforwardly, the mapping table is designed as a two-dimensional global array with a thread number, TID, cstamp, and last cstamp, which requires a lock manager for updates and read over the rows.
This degrades performance due to serialization.
Therefore, we designed a data structure that expresses the rows in a one-dimensional array
so that it can be latched with a CAS operation.
The performance was improved by avoiding serialization and cache-line conflicts for row element access.
We used {\sf RapidGC}~\cite{lim2017cicada}.
The transaction mapping table objects exploited our {\sf AssertiveVersionReuse} method, so the memory manager almost did not need to work during experiments.

{\bf TicToc:~}
We implemented all three optimization methods ({\sf NoWaitTT}, {\sf PreemptiveAbort}, {\sf TimestampHistory}).
We also improved algorithm 5 in the original paper~\cite{yu2016tictoc} by removing the redundant read timestamp updates.
This is effective because if the recorded read timestamp is the same as the previous one, it does not need to be updated.

{\bf MOCC:~}
MOCC periodically resets temperature information~\cite{wang2016mostly}, which switches many cache-lines to the invalid state simultaneously and thereby degrades performance.
In contrast, our implementation stores the epoch ID of the latest reset timing and temperature together and thereby avoids multiple resets in the same epoch.
This reduces the cost of cache-line conflicts and the number of reset operations, thereby maintaining fine control of the temperature.
We did not implement MQL since our environment had only four NUMA (non-uniform memory access) nodes, so the effect would be limited.
We instead used the reader/writer lock that was used in our 2PL implementation.

{\bf Cicada:~}
We implemented all six optimization methods:
{\sf SortWriteSetByContention},
{\sf PrecheckVersionConsistency},
{\sf AdaptiveBackoff},
{\sf EarlyAborts},
{\sf RapidGC},
and {\sf BestEffortInlining}.
Moreover, we fixed a logical bug, i.e., the incomplete version consistency check in the validation phase.
In the original paper~\cite{lim2017cicada}, only the read version is confirmed to still be the visible version.
However, whether the latest version is still the visible version needs to be confirmed as well.
The existence of a newer version in the observable range means that it should have been read and the transaction view will thus be broken.
We turned off the one-sided synchronization to improve throughput.

{\bf CCBench:~}
In the MVCC protocols (SI, ERMIA, and Cicada),
creating new versions and deleting old versions put a load on the memory manager.
We therefore developed a new optimization method dubbed, {\sf AssertiveVersionReuse} that avoids overloading the memory manager in MVCC protocols.
This method enables each worker thread to maintain a container for future version reuse.
When GC begins, the versions collected are stored in this container.
A new version is taken from this container except if it is empty.
If it is empty, a request for space is sent to the memory manager as usual.
The space needed for GC is estimated before the experiment,
and memory space is allocated in advance.
This optimization minimizes the burden on the memory manager for the MVCC protocols.
Moreover, we introduce the use of {\sf ReadPhaseExtension} to delay execution. It was inspired by the extra read process described in $\S$\ref{sec:Effect of Extra Read and Read Phase Extension Method}.
We also introduce {\sf AggressiveGC} optimization, which collects even visible versions, described in $\S$\ref{sec:discussion-towards-version-lifetime-control}.

\subsection{Configurable Workloads}
\label{Configurable Workloads}

CCBench supports the seven parameters shown in Table~\ref{tab:Experimental Parameters}.
Skew is an access pattern that follows a Zipf distribution.
Cardinality is the number of records in the database.
The payload is the size of a record (key plus value).
Transaction size is the number of operations in a transaction.
Read ratio is the ratio of reads in a transaction.
Write model is whether to perform RMW or not.
Thread count is the number of worker threads (fixed to 224 here except for reproduction).
To determine the skew, CCBench uses a fast approximation method~\cite{gray1994quickly}.
We analyzed CC protocols by using 224 threads using CCBench.
Most of the benchmarks were the YCSB workload, and some were variants of YCSB.
The analyzed parameter sets are summarized in Table \ref{tab:Experimental Parameters}.
We varied the skew, cardinality, payload, transaction size, and read ratio and fixed the thread count at 224.
The caption in the table describes the variance in parameters using $\alpha \ldots \zeta$\footnote{More spaces were explored, and they are found online \cite{thawkccdata}.}.

\begin{table}[tb]
 \begin{center}
    \begin{small}
   \caption{Analyzed parameter sets.
   ($\alpha$) Cardinality (from $10^3$ to $10^9$ records);
   ($\beta$) Cardinality ($10^3$ or $10^6$ records);
   ($\gamma$) Read ratio (from 0\% to 100\%);
   ($\delta$) Transaction size (from 10 to 100 operations);
   ($\epsilon$) Payload size (from 4 to 1000 bytes);
   ($\zeta$) Skew (from 0.6 to 0.99).
   }
 \label{tab:Experimental Parameters}
 \begin{tabular}[tb]{l|cccccc}
  & \multicolumn{2}{c}{Cache}& \multicolumn{2}{c}{Delay}& \multicolumn{1}{c}{Version}\\
  \hline
  Figure Number
  &
  $\rm{F}$\ref{fig:central_tuple1k-1g}
  & $\rm{F}$\ref{fig:Analysis of Invisible Read: YCSB, 224 Threads, Payload 4 bytes, Skew 0, 10 ops/trans.}
  & $\rm{F}$\ref{fig:Impact of transaction size: YCSB-B, 100M records, 224 threads, payload 4 bytes, skew 0.8.}
  & $\rm{F}$\ref{fig:Effect of Delay provided by Extra Read: YCSB-A, 100M records, 224 threads, skew 0.9, 10 ops/trans.}
  & $\rm{F}$\ref{fig:comp_tuple100m_skew0-099}\\
  \hline
  Skew            &0   &0   &0.8   &0.9  &$\zeta$ \\
  Cardinality     &$\alpha$ &$\beta$ &$10^8$&$10^8$&$10^8$\\
  Payload (byte)  &4   &4   &4     &$\epsilon$  &4\\
  Xact size       &10  &10  &$\delta$   &10   &10\\
  Read ratio (\%) &0,5 &$\gamma$ &50,95 &50   &50,95\\ \hline
  Thread count    &\multicolumn{5}{l}{Always 224 except from reproduction}\\
  Read modify write&\multicolumn{5}{l}{Always off except from reproduction}\\
 \end{tabular}
     \end{small}
 \end{center}
\end{table}

This paper focuses on identifying factors that significantly affect performance in a highly parallel environment. We have chosen YCSB like benchmarks because they can generate various workloads despite the simpleness of its data model which offers only read and update operations for a single table with the primary index. It is preferred that a benchmark also supports realistic workloads, which often contain insertion, deletion, range search, and secondary indexes for multiple tables. Difallah et al.~\cite{oltpbench} summarized such workloads, which included an industry-standard benchmark, TPC-C~\cite{tpcc}. CCBench currently supports only a subset of TPC-C including {\tt New-Order} and {\tt Payment} transactions, and we obtained the following results:
(1) CCBench exhibited scalability in both the thread count and the warehouse count;
(2) The majority of the execution time was for the index traversal, and its acceleration was important for high performance;
(3) Different analysis platforms exhibited different behavior even for the same workload, depending on the design and the implementation.
\ifvldb
\else
The details are described in Appendix
\ref{sec:Evaluation of a Subset of TPC-C} and
\ref{sec:Evaluation of a TPC-C Full-mix}.
CCBench will support TPC-C full-mix in future work other than Silo, which will be available at \cite{cc_format}.
\fi

\color{black} 

\section{Reproduction of Prior Work}
\label{sec:Reproduction of Prior Work}
%
%
\begin{figure*}[tb]
  \begin{tabular}{ccc}
    \begin{minipage}{0.33\hsize}
      \subcaptionbox{YCSB (Read-Only). 2 queries per transaction, skew 0.\label{fig:tuple10m_val1k_ycsbC_tps}}
      {\includegraphics[scale=1.02]{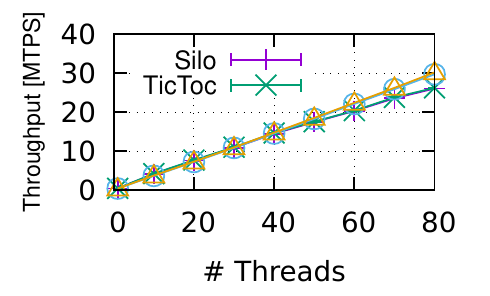}}
    \end{minipage}
    &
    \begin{minipage}{0.33\hsize}
      \subcaptionbox{YCSB (Medium Contention). 16 queries per transaction, 90\% reads, 10\% writes, skew 0.8.\label{fig:tuple10m_val1k_ycsb_medium_tps}}
      {\includegraphics[scale=1.02]{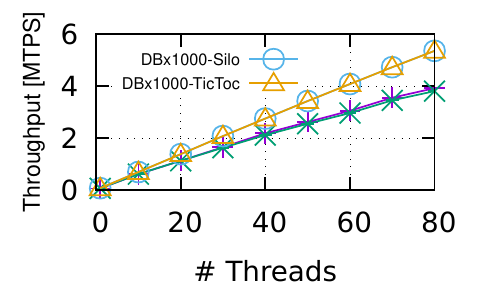}}
    \end{minipage}
    &
    \begin{minipage}{0.33\hsize}
      \subcaptionbox{YCSB (High Contention). 16 queries per transaction, 50\% reads, 50\% writes, skew 0.9.\label{fig:tuple10m_val1k_ycsb_high_tps}}
      {\includegraphics[scale=1.02]{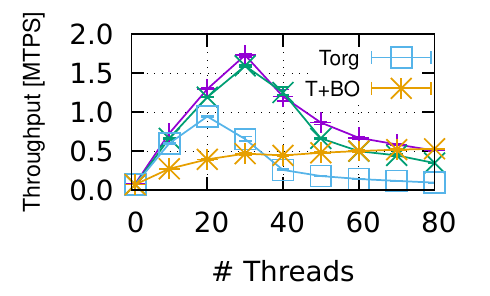}}
    \end{minipage}
  \end{tabular}
 \caption{Reproduction of TicToc experiment. 10 M records, payload 1 K bytes. Different access methods (hash for DBx1000, Masstree for CCBench). CCBench without index outperformed (39.6 Mtps) DBx1000 (30 Mtps) at 80 threads.}
\end{figure*}

%
%
\begin{figure*}[tb]
 \begin{tabular}{ccc}
  \begin{minipage}{0.3\hsize}
   \subcaptionbox{Read Only Workload.\label{fig:ycsbC_tuple50_tps}}
   {\includegraphics[scale=1.1]{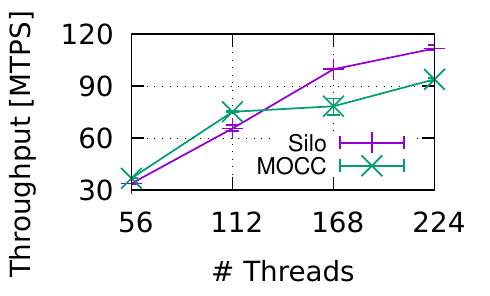}}
  \end{minipage}
  &
  \begin{minipage}{0.3\hsize}
   \subcaptionbox{Read-Write YCSB. 224 threads.\label{fig:silo-mocc_tuple50_rratio10-100_tps}}
   {\includegraphics[scale=1.1]{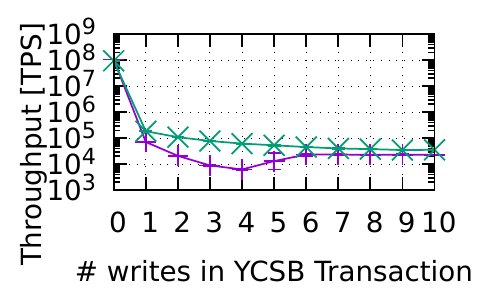}}
  \end{minipage}
  &
  \begin{minipage}{0.3\hsize}
   \subcaptionbox{Read-Write YCSB. 224 threads.\label{fig:silo-mocc_tuple50_rratio10-100_ar}}
   {\includegraphics[scale=1.1]{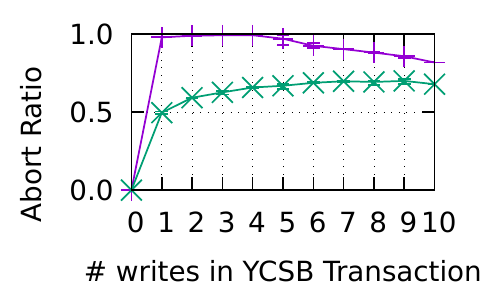}}
  \end{minipage}
 \end{tabular}
 \caption{Reproduction of MOCC experiment. YCSB, 50 records, 10 ops/trans, Skew 0, Payload 4 bytes, reads were pure read, writes were read-modify-write.}
\end{figure*}

%
%
\begin{figure*}[tb]
  \begin{tabular}{cccc}
    \begin{minipage}{0.23\hsize}
     \subcaptionbox{Write-Intensive. 16 requests/trans, skew 0.99, 50\% RMW, 50\% pure reads\label{fig:ycsbA_tuple10m_ope16_rmw_skew099_tps}.}
      {\includegraphics[scale=0.8]{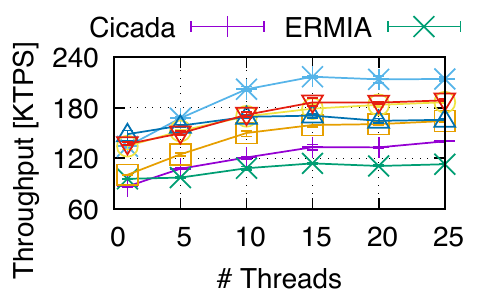}}
    \end{minipage}
    &
    \begin{minipage}{0.23\hsize}
      \subcaptionbox{Write-Intensive. 28 threads. 16 requests/trans, 50\% RMW, 50\% pure reads\label{fig:ycsbA_tuple10m_ope16_rmw_skew0-099_th28_tps}.}
      {\includegraphics[scale=0.8]{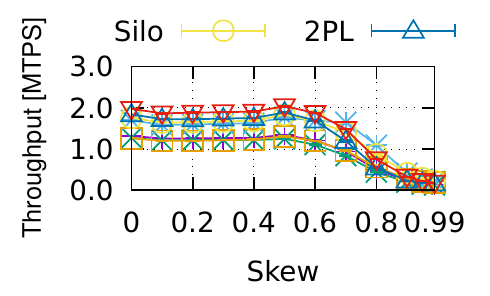}}
    \end{minipage}
    &
    \begin{minipage}{0.23\hsize}
      \subcaptionbox{Read-Intensive. 28 threads. 16 requests/trans, 5\% RMW, 95\% pure reads\label{fig:ycsbB_tuple10m_ope16_rmw_skew0-099_th28_tps}.}
      {\includegraphics[scale=0.8]{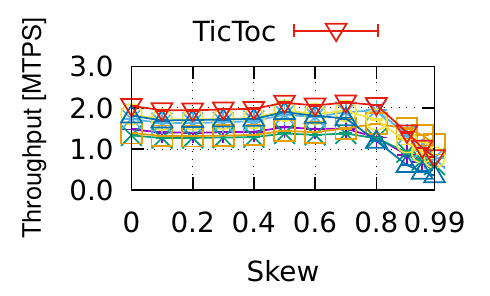}}
    \end{minipage}
   &
    \begin{minipage}{0.23\hsize}
      \subcaptionbox{Read-Intensive. 1 requests/trans, skew 0.99, 5\% RMW, 95\% pure reads\label{fig:ycsbB_tuple10m_ope1_rmw_skew099_tps}.}
      {\includegraphics[scale=0.8]{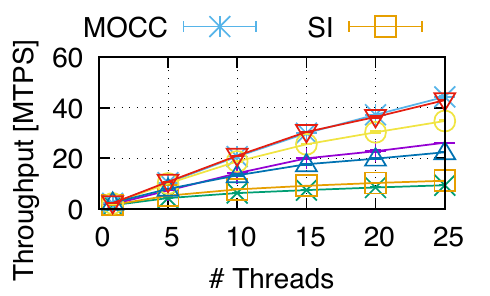}}
    \end{minipage}
  \end{tabular}
 \caption{Reproduction of Cicada experiment. 10 M records, payload 100 bytes.}
 \label{fig:Reproduction of Cicada experiment. 10M records, payload 100 bytes}
\end{figure*}

Before presenting our analysis of reproducing experiments performed in prior work,
we explain how we validated the correctness of the CCBench implementation,
so that it successfully reproduced the results of experiments performed in prior work.
We did this by evaluating DBx1000~\cite{dbx1000}, a standard analysis platform, as used by Bang et al.~\cite{10.1145/3399666.3399910}, to validate the behavior of CCBench.
For all graphs showing DBx1000 results, we set {\tt CPU\_FREQ} to 2.095 after measuring the real clock rate used for {\tt rdtscp} instruction; each plot point shows the average of ten runs. The duration of each run was 4.5 to 10 s.
As the access method, we used a hash index for DBx1000 and Masstree for CCBench.
DBx1000 does not support RMW operations in the YCSB workload, so their plots are not presented in the reproduction results for MOCC and Cicada.

%
%
\subsection{TicToc}
Our first experiment was aimed at reproducing {\rm TicToc} results.
We experimented with the settings for Figs.~6, 7, and 8 in the original paper~\cite{yu2016tictoc}.
We set the epoch interval longer than 40 ms so that cache-line conflicts for the global epoch rarely occurred in {\rm Silo}.
The experimental results are shown in Figs.~\ref{fig:tuple10m_val1k_ycsbC_tps}, \ref{fig:tuple10m_val1k_ycsb_medium_tps}, and \ref{fig:tuple10m_val1k_ycsb_high_tps}.
{\rm TicToc} uses {\sf NoWait} optimization instead of {\sf NoWaitTT}.
Note that hereafter, {\bf TicToc} means to attach {\sf NoWait} instead of {\sf NoWaitTT}.
Torg is the original TicToc with all the optimizations, including {\sf NoWaitTT}.
{\rm T+BO} is {\rm TicToc} using {\sf AdaptiveBackoff}.

Fig.~\ref{fig:tuple10m_val1k_ycsbC_tps} shows that Silo and TicToc had a comparable performance for read-only workloads, consistent with the results shown in Fig.~6 in the original paper.
Fig.~\ref{fig:tuple10m_val1k_ycsb_medium_tps} shows that Silo and TicToc both scale in performance.
This is also consistent with the results shown in Fig.~7 in the original paper.
The results for DBx1000 are also shown in the figures, and their behaviors were similar to those for CCBench.
DBx1000 outperformed CCBench, which we attribute to the difference in access methods.
A hash index was used for DBx1000 while Masstree was used for CCBench, and hash indexes are typically faster than tree indexes.
We determined that CCBench without indexing had 30\% higher performance (39.6 Mtps) than DBx1000 (30 Mtps).
As we have already stated, the effect of the access method must be considered for a fair comparison.

One finding is particularly noteworthy.
In Fig.~\ref{fig:tuple10m_val1k_ycsb_high_tps},
the behaviors of {\rm Silo} and {\rm TicToc} differ from those in Fig.~8 in the original paper~\cite{yu2016tictoc}.
The performance of {\rm TicToc} improved as the number of threads was increased, but when the number exceeded a certain value, the performance started to deteriorate due to an increase in read/write conflicts.
Therefore, we presumed that some sort of backoff was applied to {\rm TicToc} in the original paper, and attached {\sf AdaptiveBackoff} to {\rm TicToc}.
This is shown as {\tt T+BO} in Fig.~\ref{fig:tuple10m_val1k_ycsb_high_tps}.
The performance curve for {\tt T+BO} is similar to that in Fig.~9 in the original paper.
The lower performance of the original TicToc protocol shown as {\tt Torg}, is attributed to excessive delays and read validation failures.
{\sf NoWaitTT} inserts fixed delays into the validation phase, which tends to result in read validation failure.
Therefore, {\tt Torg} shows low performance in this graph, and {\rm TicToc} shows better performance due to the shorter read phase.
The results for DBx1000 are also plotted in the figures, and their behaviors are similar to our results.

Fig.~\ref{fig:tuple10m_val1k_ycsbC_tps} shows that Silo and TicToc exhibited linear scalability for a read-only workload, and Fig.~6 in the original paper shows almost linear scalability.
Fig.~\ref{fig:tuple10m_val1k_ycsb_medium_tps} above and Fig.~7(a) in the original paper show almost the same results.
In summary, CCBench tended to closely reproduce the results for {\rm TicToc}, and the results of CCBench are consistent with those of DBx1000.

%
%
\subsection{MOCC}
The second experiment was aimed at reproducing MOCC results~\cite{wang2016mostly}.
Our MOCC implementation differs from the original one in terms of temperature and MQL, as described in $\S$\ref{sec:Implementing Optimization Techniques}.
To reproduce the results in the original paper,
we experimented with the settings shown in Figs.~6 and 7 in the original paper.
The details of the workload are described in graph captions.

The results in Fig.~\ref{fig:ycsbC_tuple50_tps} appropriately reproduce those in Fig.~6 in the original paper.
The performance scales with the number of threads in the read-only workloads.
The underperformance of MOCC with more than 168 threads compared with Silo is attributed to the overhead of MOCC.
When MOCC reads a record, it accesses the corresponding temperature information and checks whether the temperature is high.

The results in Figs.~\ref{fig:silo-mocc_tuple50_rratio10-100_tps} and \ref{fig:silo-mocc_tuple50_rratio10-100_ar} closely reproduce those in Fig.~7 in the original paper. When the write ratio increased from 0\% to 10\%, the abort rate rapidly increased, and the performance deteriorated.
This is because the write operation produces a large number of read validation failures.
As shown in Fig.~\ref{fig:silo-mocc_tuple50_rratio10-100_tps}, the performance of Silo deteriorated at a write rate of 0-40\%,
while it improved at a write rate of 40-60\%.
There is thus a trade-off between reduced record-access contention due to reduced write operations and frequent read validation failures due to increased read operations.
In contrast, MOCC showed stable performance, apparently because abort is less likely to occur by temperature control.

{\rm MOCC} and {\rm Silo} exhibited scalability, as shown in Fig.~\ref{fig:ycsbC_tuple50_tps} above, while {\rm MOCC} and {\rm FOEDUS} exhibited scalability in the original paper~\cite{wang2016mostly} (in Fig.~6).
Fig.~\ref{fig:silo-mocc_tuple50_rratio10-100_tps} shows the throughputs of MOCC and Silo for
various write ratios.
They are almost the same as those in Fig.~7 in the original paper.
The abort ratios of {\rm Silo} and {\rm MOCC} shown in Fig.~\ref{fig:silo-mocc_tuple50_rratio10-100_ar} is consistent with that in Fig.~7 in the original paper.
In summary, CCBench closely reproduced the results for MOCC.

\subsection{Cicada}
\label{sec:Cicada}
Our third experiment was aimed at reproducing Cicada results.
The workload parameters were set almost the same as those in the original paper~\cite{lim2017cicada}.
The details of the workload are described in the graph caption.
The differences between this paper and the original paper are as follows:
(1) we redesigned and reimplemented all protocols;
(2) {\sf AdaptiveBackoff} was implemented in all protocols because it is easily attachable and effective for performance improvement.

Figs.~\ref{fig:ycsbA_tuple10m_ope16_rmw_skew099_tps},
\ref{fig:ycsbA_tuple10m_ope16_rmw_skew0-099_th28_tps},
\ref{fig:ycsbB_tuple10m_ope16_rmw_skew0-099_th28_tps},
and
\ref{fig:ycsbB_tuple10m_ope1_rmw_skew099_tps}
present the experimental results reproducing those in Figs.~6a, 6b, 6c, and 7 in the original paper, respectively.
In Figs.~\ref{fig:ycsbA_tuple10m_ope16_rmw_skew0-099_th28_tps},
\ref{fig:ycsbB_tuple10m_ope16_rmw_skew0-099_th28_tps},
and \ref{fig:ycsbB_tuple10m_ope1_rmw_skew099_tps}, the tendencies in our results for Silo, TicToc, MOCC, ERMIA, and 2PL are
the same as those in the corresponding figures in the original paper.
In contrast, the results for Cicada differ.
Fig.~\ref{fig:ycsbB_tuple10m_ope1_rmw_skew099_tps} shows that Cicada had worse performance than Silo and TicToc for a read-intensive workload, and this is inconsistent with the results shown in Fig.~7 in the original paper.
We discuss this inconsistency in {\bf Insight 5} ($\S$\ref{sec:problem-many-visible-versions}).

\section{Analysis of Cache}
\label{sec:Cache Effect}

Here, we discuss two effects of the cache.
{\bf Cache-line conflict} occurs when many worker threads access to the same cache-line with some writes (e.g., accessing a single shared counter).
The cache-line becomes occupied by a single thread, and the other threads are internally blocked.
{\bf Cache-line replacement} occurs when transactions access a large amount of data (e.g., high cardinality, low skew, or large payload).
The data accesses tend to replace the contents of the L1/L2/L3 cache.

\subsection{Cache-Line Conflict}
\label{sec:changing_database_size}
\begin{figure*}[tb]
 \begin{tabular}{cccc}
  \hspace{-0.4cm}
  \begin{minipage}{0.25\hsize}
   \subcaptionbox{YCSB-B.\label{fig:ycsbB_tuple1k-1g}}
   {\includegraphics[scale=0.8]{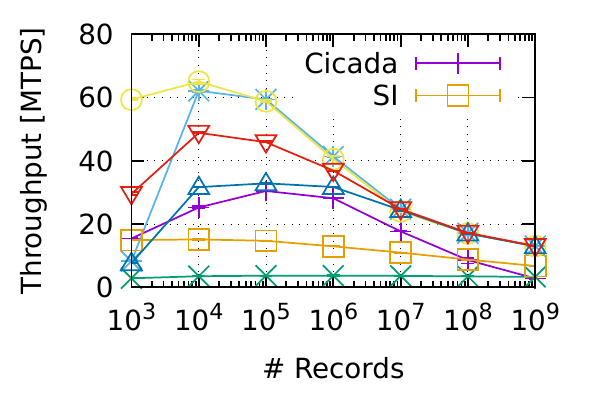}}
  \end{minipage}
  &
  \hspace{-0.5cm}
  \begin{minipage}{0.25\hsize}
   \subcaptionbox{YCSB-C.\label{fig:ycsbC_tuple1k-1g}}
   {\includegraphics[scale=0.8]{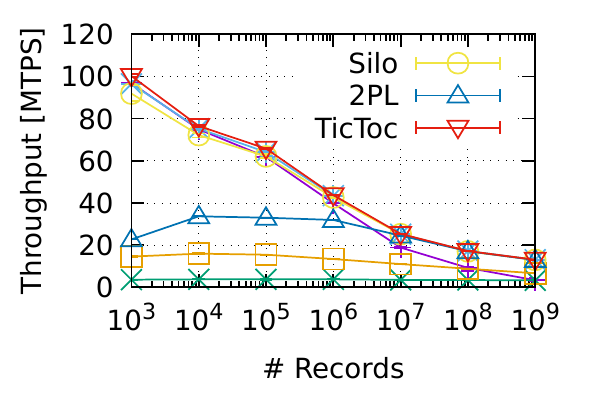}}
  \end{minipage}
  &
  \hspace{-0.5cm}
  \begin{minipage}{0.25\hsize}
   \subcaptionbox{YCSB-B.\label{fig:ycsbB_tuple1k-1g_ar}}
   {\includegraphics[scale=0.8]{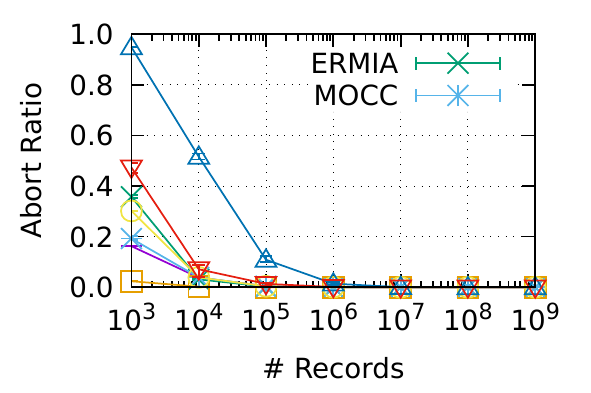}}
  \end{minipage}
  &
  \hspace{-0.5cm}
  \begin{minipage}{0.25\hsize}
   \subcaptionbox{YCSB-C.\label{fig:ycsbC_tuple1k-1g_ca}}
   {\includegraphics[scale=0.8]{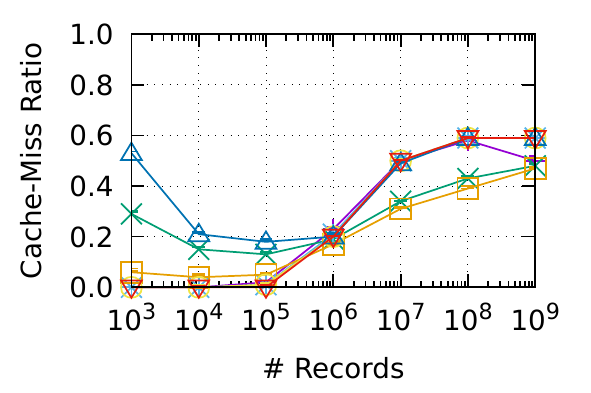}}
  \end{minipage}
 \end{tabular}
 \caption{Varying cardinality: payload 4 bytes, 224 threads, skew 0, 10 ops/trans.}
 \label{fig:central_tuple1k-1g}
\end{figure*}

\begin{figure}[t]
  \begin{tabular}{cc}
   \begin{minipage}{0.45\hsize}
      \subcaptionbox{Throughput.\label{fig:fetch_add_test_tps}}
      {\includegraphics[scale=0.8]{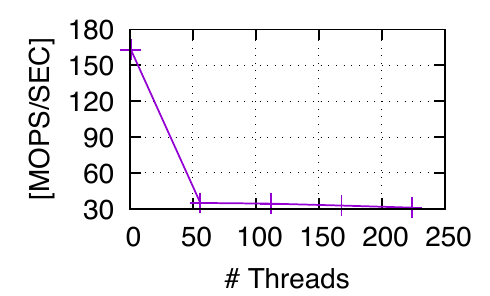}}
   \end{minipage}
    &
    \begin{minipage}{0.45\hsize}
      \subcaptionbox{Cache miss ratio.\label{fig:fetch_add_test_cm}}
      {\includegraphics[scale=0.8]{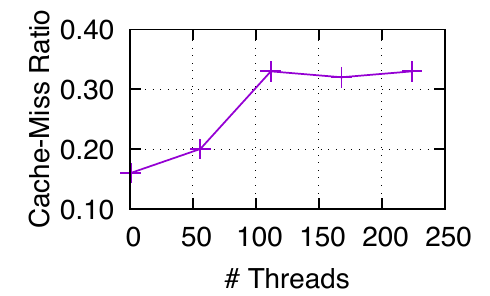}}
    \end{minipage}
  \end{tabular}
  \caption{Scalability of fetch\_add.}
  \label{fig:fetch_add}
\end{figure}

Some protocols use centralized ordering (e.g., a shared counter)~\cite{jung2013performance,kim2016ermia} while others use decentralized ordering~
\cite{
  lim2017cicada,
  tu2013speedy,
  wang2016mostly,
  yu2016tictoc}.
We first analyzed the negative effect of centralized ordering.
The results are shown in Fig.~\ref{fig:central_tuple1k-1g}.
Since the results for YCSB-A have the same tendency as those for YCSB-B, they are omitted.
Fig.~\ref{fig:central_tuple1k-1g} shows that the throughputs of ERMIA and SI did not scale although their scheduling spaces are wider than those of single-version concurrency control (SVCC) serializable protocols.
This is because both protocols depend on a centralized data structure, which causes cache-line conflicts, which in turn degrade performance.
This is consistent with previous findings~\cite{tu2013speedy, wang2016mostly}.

To investigate the cause of the inefficiency, we measured the throughput of {\tt fetch\_add}.
The results are shown in Figs.~\ref{fig:fetch_add}.
The L3 cache miss rate was measured using the Linux perf tool.
Fig.~\ref{fig:fetch_add_test_tps} shows that one thread exhibited the best performance.
This is because frequent reads/writes to the same memory address by multiple threads cause many cache-line conflicts.
Fig.~\ref{fig:fetch_add_test_cm} shows that the L3 cache miss rate increased with the number of threads.
%
Cache-line conflicts often result in L3 cache misses and longer communication delays because the CPU core uses a cache-line, and the other cores running in different CPU sockets also require the line.
In the experiment setting, 56 threads used two sockets, and 112 or more threads used four sockets.
Thus, a higher L3 cache miss rate ($\geq$ 30\%) indicates communication among sockets, which greatly degrades the performance of {\tt fetch\_add}.
This means that centralized ordering should be avoided.

\subsection{Cache-Line Replacement}
\label{sec:Cache Miss Mitigation}
We discuss the results for YCSB-B.
In Figs.~\ref{fig:ycsbB_tuple1k-1g} and \ref{fig:ycsbB_tuple1k-1g_ar}, as cardinality increased,
(1) the throughput first improved and then deteriorated, and
(2) the abort ratio monotonically improved.
When cardinality was quite low, the footprint of accessible records in database was small so that all of the records fit inside the L3 cache.
The contention is more likely with L1 or L2 cache.
As cardinality increases, such contention is alleviated since the number of accessible records increases, and the performance improves.
It is likely that with more cardinality, the total number of records starts to overflow L3 cache.
This results in L3 cache misses, and consecutive accesses to remote socket cache or DRAM degrades the performance.
%

We investigated the strong effect of the L3 cache miss ratio.
As shown in Fig.~\ref{fig:ycsbC_tuple1k-1g_ca},
when cardinality was no more than $10^5$, the L3 cache miss ratios for Silo, TicToc, and Cicada were almost zero.
This is because the L3 cache size was 38.5 MB and the size of a record including the header was 64 bytes (64 $\times 10^5 < 38.5 \times 10^6$).
However, as cardinality increased, their throughputs decreased linearly, as shown in Fig.~\ref{fig:ycsbC_tuple1k-1g}.
This is attributed to L1/L2 cache misses.
When cardinality was more than $10^5$, we observed that L3 cache misses affected performance even for read-only workloads.

The negative effect of read contention is shown in Figs.~1 and 6 in the original paper~\cite{wang2016mostly}.
The cache-line conflicts that occurred due to read lock contention degraded performance, and the difference in performance between OCC and 2PL rapidly increased as the number of threads increased.
However, there is no mention in the paper that the greater the number of L3 cache misses for a read-only workload, the smaller the performance difference between OCC and 2PL.
As shown in Fig.~\ref{fig:ycsbC_tuple1k-1g}, when cardinality was no less than $10^7$, Silo, TicToc, and 2PL had almost the same throughput.
Moreover, their L3 cache miss ratios were almost the same, as shown in Fig.~\ref{fig:ycsbC_tuple1k-1g_ca}.
For read-only workloads, the differences in performance among protocols converged due to the L3 cache misses when the database size was large.
This has not been reported so far.

\underline{\bf Insight 1:}~
Even if the entire database is within L3 cache,
as cardinality increases,
(1) OCC for read-intensive workloads improves due to a decrease in L1/L2 cache-line conflicts, and
(2) OCC for read-only workloads deteriorates due to an increase in L1/L2 cache-line replacements.
If the entire database slightly overflows the L3 cache, the performances of the protocols diverge.
If the size of the entire database is much larger than the L3 cache, the performances of Silo, TicToc, MOCC, and 2PL converge due to frequent L3 cache-line replacements.
%

\begin{figure*}[tb]
  \begin{tabular}{ccc}
    \begin{minipage}{0.33\hsize}
      \subcaptionbox{Throughput.\label{fig:ycsb_rratio10-90_tps}}
      {\includegraphics[scale=1]{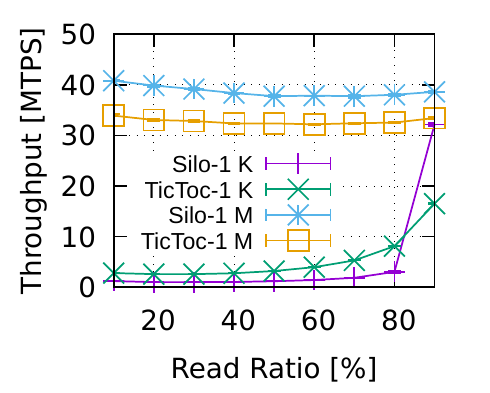}}
    \end{minipage}
    &
    \begin{minipage}{0.33\hsize}
      \subcaptionbox{Abort ratio.\label{fig:ycsb_rratio10-90_ar}}
      {\includegraphics[scale=1]{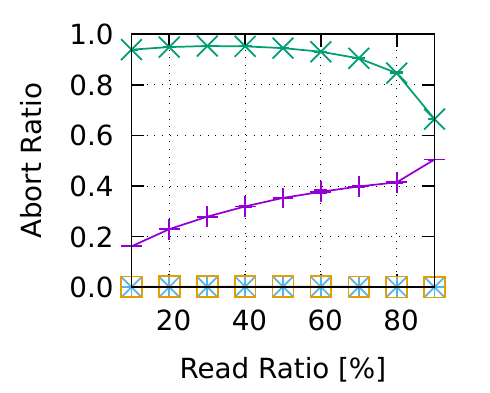}}
    \end{minipage}
    &
    \begin{minipage}{0.33\hsize}
      \subcaptionbox{Cache-miss ratio.\label{fig:ycsb_rratio10-90_ca}}
      {\includegraphics[scale=1]{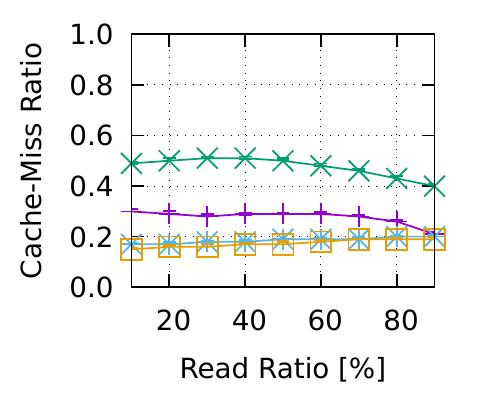}}
    \end{minipage}
  \end{tabular}
  \caption{Analysis of {\sf InvisibleReads}. 224 threads, payload 4 bytes, skew 0, 10 ops/trans. cardinality $10^3$ or $10^6$ records.}
  \label{fig:Analysis of Invisible Read: YCSB, 224 Threads, Payload 4 bytes, Skew 0, 10 ops/trans.}
\end{figure*}

\subsection{Effect of Invisible Reads}
\label{sec:Effect of Invisible Read}
Updating shared data among threads running in different CPU sockets trigger L3 cache misses. It is mainly due to the cache-line conflicts, leading to performance degradation.
The read operation in Silo does not update the corresponding metadata, which can prevent cache-line conflicts.
This read method is referred to as {\sf InvisibleReads}~\cite{marathe2004design}.
It effectively improves performance of protocols for read-intensive workloads; however, its behaviors have not been explored yet.
The read operations of TicToc typically update the corresponding metadata (i.e., read timestamp), so they are not {\sf InvisibleReads}.

%
To investigate the details of {\sf InvisibleReads}, we measured the performance of Silo and TicToc for various read ratios and cardinalities.
The results of experiments are shown in Figs.~\ref{fig:ycsb_rratio10-90_tps}-\ref{fig:ycsb_rratio10-90_ca}.
Fig.~\ref{fig:ycsb_rratio10-90_tps} shows that TicToc-1 K (where 1 K means 1000 records) outperformed Silo-1 K for low cardinality and write-intensive workloads.
%
%
The abort ratio of TicToc-1 K was much worse than that of Silo-1 K, as shown in Fig.~\ref{fig:ycsb_rratio10-90_ar}.
As shown in Fig.~\ref{fig:ycsb_rratio10-90_tps}, in the read-most (90\%) case,
Silo-1 K outperformed TicToc-1 K, as evidenced by the rapid increase in the abort ratio (Fig.~\ref{fig:ycsb_rratio10-90_ar}).
The {\sf InvisibleReads} method contributed to the performance improvement in this case.
For high cardinality workloads, Silo-1 M (where 1 M means 1 million records) always outperformed TicToc-1 M.
The abort ratios of Silo-1 M and TicToc-1 M were almost zero, as shown in Fig.~\ref{fig:ycsb_rratio10-90_ar}.
This result seems mysterious. The abort ratios for both Silo and TicToc were almost the same, and their L3 cache miss ratios were almost the same, as shown in Fig.~\ref{fig:ycsb_rratio10-90_ca}. However, Silo-1 M outperformed TicToc-1 M.
This is because dealing with contention resolution processing in TicToc
requires three functions: timestamp history management, early abort judgment, and read timestamp update.
These functions require executing additional instructions and thus degrade performance.
If {\sf InvisibleReads} is woven into the process, a protocol has difficulty exploiting them.

\underline{\bf Insight 2:}~
For a detailed analysis of protocols, contention should be carefully considered.
The occurrence of contention depends not only on the read ratio but also on cardinality and skew.
We observed that Silo outperformed TicToc not only for read-only workloads but also for write-intensive workloads.
This fact has not been reported in previous studies~\cite{yu2016tictoc, lim2017cicada, trireme}.
%
These results are due to the use of {\sf InvisibleReads} method, which prevents cache misses in low contention cases. The efficiency of {\sf InvisibleReads} is provided by its simple design that does not require computations for conflict resolution or anti-dependency tracking.

\section{Analysis of Delay}
\label{sec:Delay Control}

\begin{figure}[t]
  \begin{tabular}{cc}
   \hspace{-0.3cm}
   \begin{minipage}{0.45\hsize}
     \begin{flushleft}
      \subcaptionbox{Throughput.\label{fig:ycsbB_tuple100m_skew06-085_tps}}
      {\includegraphics[scale=0.9]{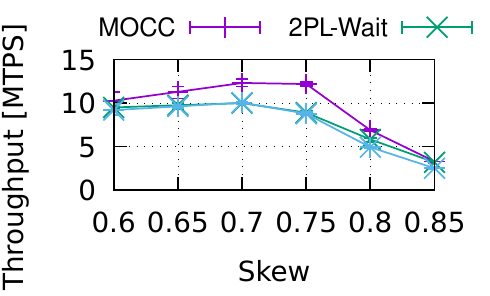}}
     \end{flushleft}
   \end{minipage}
    &
    \begin{minipage}{0.45\hsize}
      \subcaptionbox{Abort ratio.\label{fig:ycsbB_tuple100m_skew06-085_ar}}
      {\includegraphics[scale=0.9]{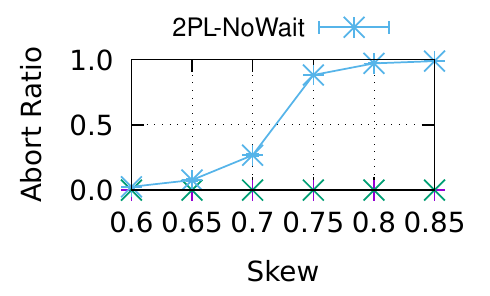}}
    \end{minipage}
  \end{tabular}
  \caption{{\sf Wait} or {\sf NoWait}: YCSB-B, 100 M records, 224 threads, payload 4 bytes, 10 ops/trans.}
  \label{fig:skew_on_large_records}
\end{figure}

%
%
\begin{figure}[t]
  \begin{tabular}{cc}
   \hspace{-0.4cm}
   \begin{minipage}{0.45\hsize}
     \subcaptionbox{Throughput.\label{fig:ycsbB_tuple100m_skew08_ope10-100_tps}}
     {\includegraphics[scale=0.9]{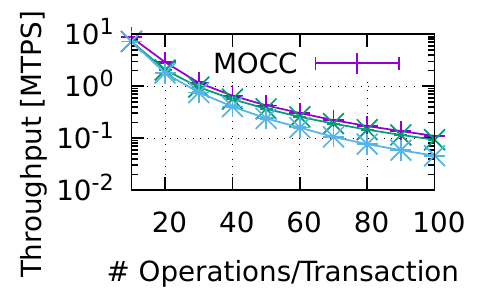}}
   \end{minipage}
    &
    \begin{minipage}{0.45\hsize}
     \subcaptionbox{Abort ratio.\label{fig:ycsbB_tuple100m_skew08_ope10-100_ar}}
     {\includegraphics[scale=0.9]{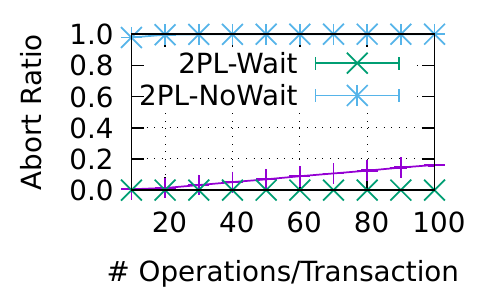}}
    \end{minipage}
  \end{tabular}
 \caption{Impact of transaction size: YCSB-B, 100 M records, 224 threads, payload 4 bytes, skew 0.8.}
 \label{fig:Impact of transaction size: YCSB-B, 100M records, 224 threads, payload 4 bytes, skew 0.8.}
\end{figure}

%
%
\begin{figure*}[t]
  \begin{tabular}{ccc}
    \begin{minipage}{0.3\hsize}
     \subcaptionbox{Throughput.\label{fig:ycsbA_tuple100m_skew09_val4-1k_tps}}
      {\includegraphics[scale=1.15]{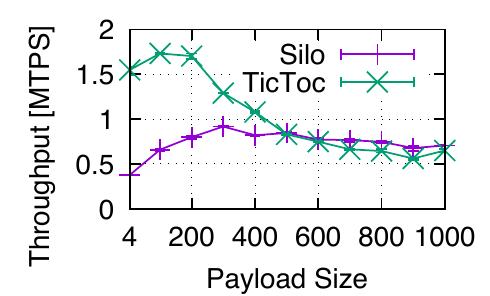}}
    \end{minipage}
   &
    \begin{minipage}{0.3\hsize}
     \subcaptionbox{Extra reads.\label{fig:ycsbA_tuple100m_skew09_val4-1k_er}}
     {\includegraphics[scale=1.15]{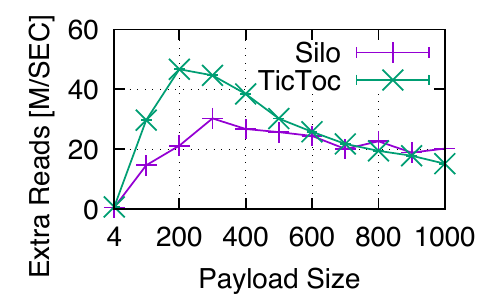}}
    \end{minipage}
   &
    \begin{minipage}{0.3\hsize}
     \subcaptionbox{{\sf ReadPhaseExtension.}\label{fig:ycsbA_tuple100m_skew09_slprp0-1000_tps}}
     {\includegraphics[scale=1.15]{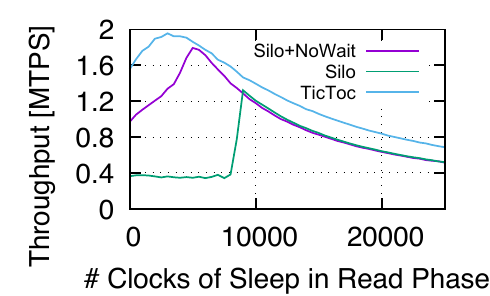}}
    \end{minipage}
  \end{tabular}
 \caption{Effect of delay provided by extra read: YCSB-A, 100 M records, 224 threads, skew 0.9, 10 ops/trans.}
 \label{fig:Effect of Delay provided by Extra Read: YCSB-A, 100M records, 224 threads, skew 0.9, 10 ops/trans.}
\end{figure*}

\subsection{Effect of NoWait}
\label{sec:Confirmation of Effect of NoWait}

A method for mitigating performance degradation caused by conflicts is to place an additional artificial delay before a retry, such as with {\sf NoWaitTT}~\cite{yu2016tictoc} and {\sf AdaptiveBackoff}~\cite{lim2017cicada}.
In contrast to waiting for a lock release, the {\sf NoWait} method immediately releases all of acquired locks when a conflict is detected and immediately retries. 
The question then in dealing with contention is whether to choose
{\sf NoWait} or {\sf Wait}.

To answer this question, we compared the performance of three methods for various degrees of skew: 2PL-Wait, 2PL-NoWait, and MOCC (mix of OCC and 2PL-Wait).
All operations of each transaction for 2PL-Wait were sorted in advance to avoid deadlock.
As shown in Fig.~\ref{fig:ycsbB_tuple100m_skew06-085_tps},
the performance of all three protocols started to degrade at a skew of 0.7 due to a large number of conflicts.
%
%
%
MOCC demonstrated excellent performance, as shown in Fig.~\ref{fig:ycsbB_tuple100m_skew06-085_tps}. This is because MOCC manages fewer lock objects for workloads with less skew.
%
2PL-Wait and 2PL-NoWait showed similar performances at skews of 0.60-0.85, so the cost of abort-retry and the cost of waiting for lock were similar.
So, we need more contention to answer the question.

To provide more contention, we increased the transaction size.
The cost of aborts should be smaller for short transactions and larger for long ones.
Therefore, with long transactions, the abort-retry cost is higher and may exceed the lock waiting cost.
%
%
Fig.~\ref{fig:skew_on_large_records} shows that the lock waiting cost of {\sf Wait} and the abort-retry cost of {\sf NoWait} were similar.
Fig.~\ref{fig:ycsbB_tuple100m_skew08_ope10-100_tps} shows that {\sf Wait} began to outperform {\sf NoWait} as the transaction size increased.
%
%
%
In contrast, Silo-NoWait outperformed Silo(-Wait), as shown in
Fig.~\ref{fig:Effect of Delay provided by Extra Read: YCSB-A, 100M records, 224 threads, skew 0.9, 10 ops/trans.} (described in $\S$\ref{sec:Effect of Extra Read and Read Phase Extension Method}).
The answer then is that it depends on the situations.
Therefore, the control of waiting time is important for efficiency.

\underline{\bf Insight 3:}
One cannot give a general statement about which is better, {\sf NoWait} or {\sf Wait} since it depends on the situations.
As shown in Fig.~\ref{fig:ycsbB_tuple100m_skew08_ope10-100_tps}, {\sf Wait} begins to outperform {\sf NoWait} as the transaction size increases.
On the other hand, as shown in Fig.~\ref{fig:ycsbA_tuple100m_skew09_slprp0-1000_tps}, Silo with {\sf NoWait} comes to outperform the original Silo.
This suggests the importance of adaptive behavior.
Although adaptive concurrency controls have been studied~\cite{wang2016mostly,AOCC_Guo,abortML,PelotonConcept},
their importance has not been discussed in depth regarding optimizations except for backoff~\cite{lim2017cicada}.
%
Studies on adaptive optimization methods remain on the frontier.

%
%
\subsection{Effect of Payload Size}
\label{sec:Effect of Extra Read and Read Phase Extension Method}
We analyzed the effect of payload size, which had not previously been done, to the best of our knowledge.
Fig.~\ref{fig:ycsbA_tuple100m_skew09_val4-1k_tps}
shows the relationship between throughput and payload size for Silo and TicToc.
Throughput initially increased with payload size, and then started to decrease at a certain point.
An increase in payload size would likely degrade OCC performance, which is consistent with the throughput decrease shown in the right half of the graph.
However, the throughput increase shown in the left half of the graph is counter-intuitive.

We hypothesized that this increase was due to the delay caused by the extra reads process, where two or more reads were performed if there was an interruption in the concurrent update transactions.
An increase in payload size lengthens the time to read the payload.
This increases the probability of an interruption due to an update transaction,
which increases the number of extra reads.
This behavior extends the read phase and reduces the number of transactions that run through the validation phase in parallel.
As the number of committed update transactions decreases, the number of read validation failures also decreases, which leads to throughput improvement.
Delaying the retry of atomic reads reduces the number of concurrent worker threads involved in the validation phase for the entire system, which also reduces contention.
%

Besides {\sf Wait}, {\sf NoWait}, and {\sf AdaptiveBackoff},
we present a novel delay-related optimization method,
{\sf ReadPhaseExtension}, which was inspired by the positive effect of extra reads.
Comparing Figs.~\ref{fig:ycsbA_tuple100m_skew09_val4-1k_tps} and \ref{fig:ycsbA_tuple100m_skew09_val4-1k_er} reveal that the curves are similar.
This indicates that throughput and the number of extra reads are correlated.
If this is correct, adding an artificial delay into the read phase should produce similar results.
We conducted such an experiment for a payload size of 4 bytes.
The result in Fig.~\ref{fig:ycsbA_tuple100m_skew09_slprp0-1000_tps} shows that a certain amount of additional delay (less than 9000 clocks) improved performance. This effect is similar to that of the extra read process.
We refer to this artificial delay as {\sf ReadPhaseExtension} and define it as a new optimization method.
{\sf ReadPhaseExtension} is configured by setting the delay on the basis of local conflict information.
This optimization method can exploit information on the access conflicts for each record during the read phase whereas {\sf AdaptiveBackoff} uses only global information across all worker threads.

\underline{\bf Insight 4:}~
The extra read process plays a key role in the performance of OCC protocols.
It is known that the contention regulation caused by delay can contribute to performance improvement depending on the situation.
A remarkable finding here is that the {\sf ReadPhaseExtension} inspired by the extra read process can also improve performance.
{\sf ReadPhaseExtension} differs from {\sf NoWaitTT} since it can exploits information on conflicting records inside transactions to adjust the delay whereas the delay in {\sf NoWaitTT} is fixed.
Such additional delay in the read phase and the use of thread-local conflict information combine to create a unique optimization method.
%

\section{Analysis of Version Lifetime}
\label{sec:version-lifetime-control}
\begin{figure*}[tb]
\begin{tabular}{cccc}
\begin{minipage}{0.23\hsize}
\subcaptionbox{YCSB-A.\label{fig:comp_ycsbA_tuple100m_skew0-099_tps}}
{\includegraphics[scale=0.9]{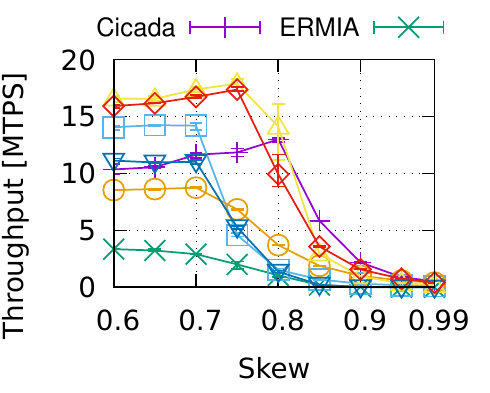}}
 \end{minipage}
&
\begin{minipage}{0.23\hsize}
\subcaptionbox{YCSB-A (high skew).\label{fig:comp_ycsbA_tuple100m_skew0-099_ar}}
{\includegraphics[scale=0.9]{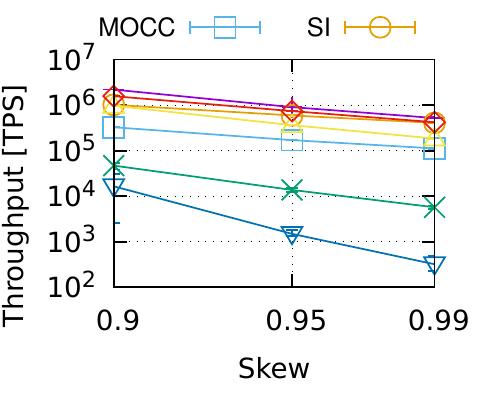}}
\end{minipage}
&
\begin{minipage}{0.23\hsize}
\subcaptionbox{YCSB-B.\label{fig:comp_ycsbB_tuple100m_skew0-099_tps}}
{\includegraphics[scale=0.9]{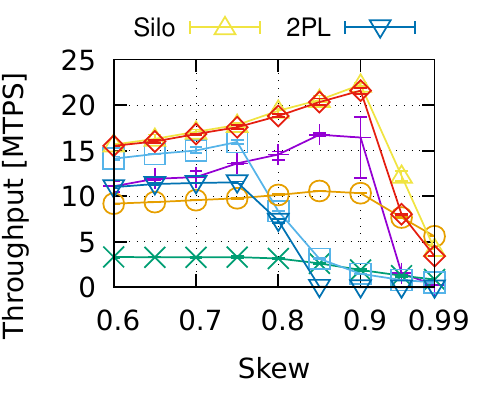}}
\end{minipage}
&
\begin{minipage}{0.23\hsize}
\subcaptionbox{YCSB-B (high skew).\label{fig:comp_ycsbB_tuple100m_skew0-099_ar}}
{\includegraphics[scale=0.9]{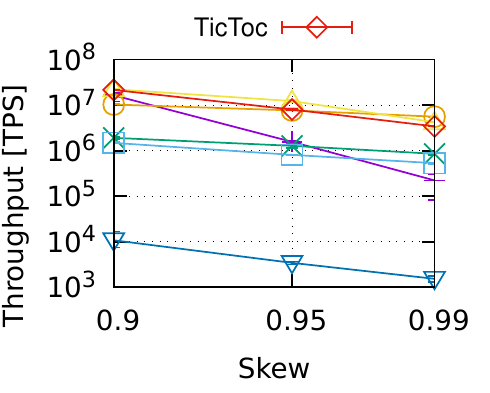}}
\end{minipage}
\end{tabular}
\caption{Effect of workload skew: 100 M records, 224 threads, payload 4 bytes, 10 ops/trans.}
\label{fig:comp_tuple100m_skew0-099}
	\end{figure*}

\begin{figure}[tb]
\begin{tabular}{cc}
\hspace{-0.3cm}
 \begin{minipage}{0.45\hsize}
\subcaptionbox{YCSB-A.\label{fig:comp_cicada-silo_ycsbA_tuple100m_gci10us_224th_main-latency}}
{\includegraphics[scale=0.9]{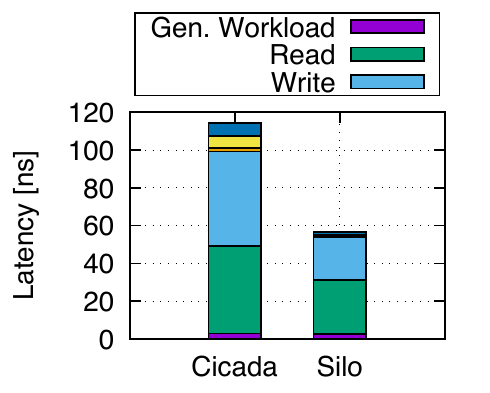}}
 \end{minipage}
&
 \begin{minipage}{0.45\hsize}
\subcaptionbox{YCSB-B.\label{fig:comp_cicada-silo_ycsbB_tuple100m_gci10us_224th_main-latency}}
{\includegraphics[scale=0.9]{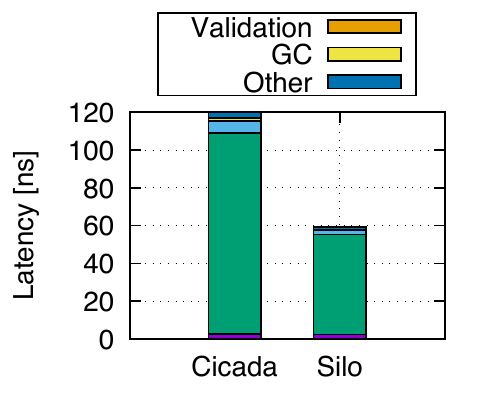}}
 \end{minipage}
\end{tabular}
\caption{Latency breakdown: skew 0, 100 M records, 224 threads, payload 4 bytes, 10 ops/trans.}
\label{fig:latency_breakdown_tuple100m_skew0}
\end{figure}

\begin{figure}[tb]
\begin{tabular}{cc}
\hspace{-0.3cm}
 \begin{minipage}{0.45\hsize}
\subcaptionbox{YCSB-A.\label{fig:comp_cicada-silo-part_ycsbA_tuple100m_gci10us_224th_main-latency}}
{\includegraphics[scale=0.9]{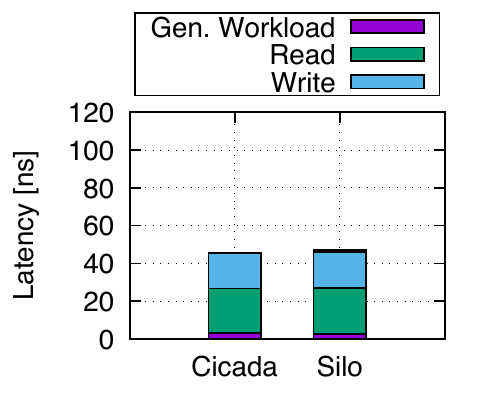}}
 \end{minipage}
&
\begin{minipage}{0.45\hsize}
\subcaptionbox{YCSB-B.\label{fig:comp_cicada-silo-part_ycsbB_tuple100m_gci10us_224th_main-latency}}
{\includegraphics[scale=0.9]{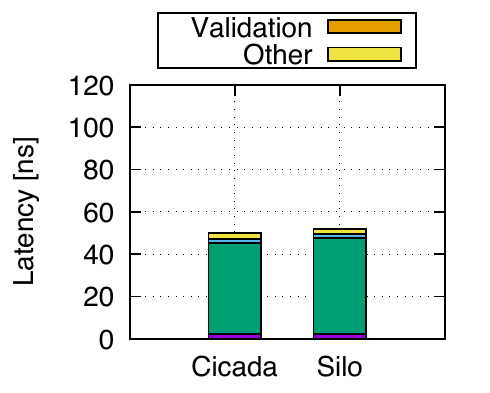}}
\end{minipage}
\end{tabular}
 \caption{Latency breakdown of Cicada-SV and Silo for partitioned workload. Other settings were the same as for Fig. 12.}
\label{fig:latency_breakdown_tuple100m_skew0_cicada-sv}
\end{figure}

%
%
\begin{figure}[t]
  \centering
  \includegraphics[scale=1.0]{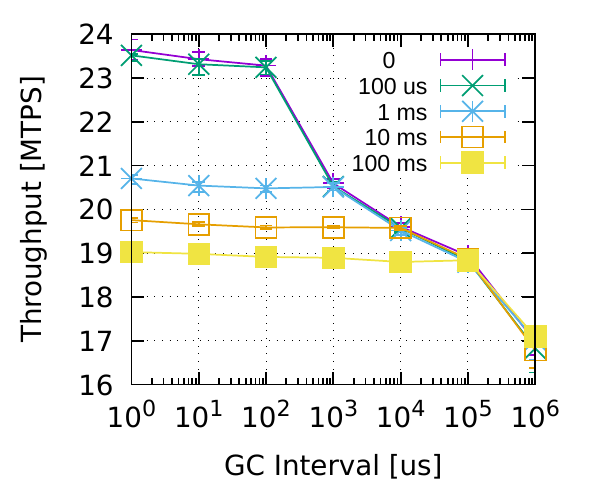}
  \caption{Analysis of RapidGC: YCSB-A, skew 0, 224 threads, 1 M records, payload 4 bytes, 10 ops/trans. Inserted delays for long transactions: 0 to 10 ms.}
  \label{fig:comp_cicada-ninline_ycsbA_tuple1m_gci1us-1s_tps}
\end{figure}

\subsection{Determining Version Overhead}
\label{sec:problem-many-visible-versions}


The life of a version begins when a corresponding write operation creates it.
The version state is called {\bf visible} during the period when other transactions can read it.
Otherwise, the version state is called {\bf non-visible}.
Recent CC protocols typically make the version visible at the pre-commit phase~\cite{aether,tu2013speedy}.
After a certain period, the life of the version ends, and it is made non-visible.
SVCC protocols typically make a version by overwriting its previous version,
with the former becoming visible and the latter becoming non-visible at the same time.
MVCC protocols typically make a version on a different memory fragment from other visible versions of the same record. Therefore, their life does not end unless GC conducts its work.

The performance of MVCC is thought to suffer from the existence of many visible versions~\cite{10.14778/2095686.2095689,10.1145/2723372.2749436}.
They lead to a significant burden due to memory exhaustion or an increase in cache-line replacements for traversal or installation.
MVCC protocols can provide higher performance than SVCC ones due to their larger scheduling spaces.
It should be noted that architectural factors sometimes negate the benefits, as mentioned in $\S$\ref{sec:changing_database_size}.
We analyzed this situation in depth.
Note that the cost for allocating new pages is negligible in the results below since our {\sf AssertiveVersionReuse} ($\S$\ref{sec:A Platform for Fair Comparison}) conducts the space allocation beforehand.

MVCC protocols often exhibit high performance compared with SVCC protocols for high contention workloads.
To begin, we measured the throughput for various degrees of skew (0 to 0.99), as shown in Fig.~\ref{fig:comp_tuple100m_skew0-099}.
Here we focus on the difference between Cicada (MVCC) and Silo (SVCC).
In Figs.~\ref{fig:comp_ycsbA_tuple100m_skew0-099_tps} and \ref{fig:comp_ycsbA_tuple100m_skew0-099_ar} for YCSB-A, Cicada outperformed Silo when the skew was higher than 0.8.
The larger the payload, the greater the superiority of Cicada.
This is attributed to Cicada having larger scheduling spaces than Silo;
nevertheless, these results confirm that the performance of MVCC suffers from having many versions~\cite{10.14778/2095686.2095689,10.1145/2723372.2749436}.
In Figs.~\ref{fig:comp_ycsbB_tuple100m_skew0-099_tps} and \ref{fig:comp_ycsbB_tuple100m_skew0-099_ar}, Cicada underperformed Silo, TicToc, and MOCC on YCSB-B.

To clarify the reason, we analyzed the characteristics of the versions.
Larger scheduling spaces are supported by many versions, and they can be an overhead.
Such overhead in modern protocols remains to be analyzed.
We show the latency breakdown with a skew of 0, where MVCC is at a disadvantage, in Fig.~\ref{fig:latency_breakdown_tuple100m_skew0}.
In Figs.~\ref{fig:comp_cicada-silo_ycsbA_tuple100m_gci10us_224th_main-latency} and ~\ref{fig:comp_cicada-silo_ycsbB_tuple100m_gci10us_224th_main-latency},
we can see that the major overhead of Cicada lies in the read and write operations rather than validation or GC.
To clarify the reasons for the overhead, we designed an SVCC version of Cicada, Cicada-SV, that overwrites the inline versions of Cicada.
This eliminates the need to allocate new areas, so Cicada-SV performs like an SVCC protocol.
We prepared a special workload in which transactions were completely partitioned to avoid contention because Cicada-SV does not cope well with contention.
The results are shown in Fig.~\ref{fig:latency_breakdown_tuple100m_skew0_cicada-sv}.
The latencies of Cicada-SV and Silo were almost the same.
We attribute the overhead shown by Cicada in Figs.~\ref{fig:comp_cicada-silo_ycsbA_tuple100m_gci10us_224th_main-latency} and \ref{fig:comp_cicada-silo_ycsbB_tuple100m_gci10us_224th_main-latency} to the cost of version chain traversals.

Since having many visible versions degrades performance, reducing the number should improve performance.
{\sf RapidGC} optimization effectively mitigates this problem.
To evaluate its efficiency, we varied the interval setting and measured the throughput of Cicada.
The legend {\tt 0} and {\tt 100 us} in Fig.~\ref{fig:comp_cicada-ninline_ycsbA_tuple1m_gci1us-1s_tps} show that high throughputs were obtained when the interval was no more than 100 us, as reported in prior studies~\cite{lim2017cicada,wu2017empirical}.
This is because the number of visible versions was small, and most of them were kept in cache.
So, can {\sf RapidGC} with its short intervals perform efficiently for any workload? The answer is no. We explain the reason why in $\S$\ref{sec:limitation-of-current-approaches}.


\underline{\bf Insight 5:~}
The overhead of multi-version management is not negligible.
Silo and TicToc outperformed Cicada in high-skew (0.99), read-intensive (YCSB-B), non-RMW, high cardinality (100 M records) cases.
A previous study~\cite{lim2017cicada} found that all three exhibited a similar performance for a similar workload (skew 0.99, read 95\%, RMW, 10 M records, payload 100 byte), as shown in Fig.~7 in the original paper.
This inconsistency is due to the use of different platforms in the previous study.
Using a single platform, we observed the difference and found that the version management cost of Cicada is not negligible even for a low contention case, 
as shown in Figs.~\ref{fig:comp_cicada-silo_ycsbA_tuple100m_gci10us_224th_main-latency} and \ref{fig:comp_cicada-silo_ycsbB_tuple100m_gci10us_224th_main-latency}.
It is difficult to obtain precise knowledge using different reference implementations or platforms, and the deep analysis of CC protocols must be done on the same platform.

\subsection{Limit of Current Approach}
\label{sec:limitation-of-current-approaches}
We investigated the behavior of {\sf RapidGC} using the same workloads as in our third experiment with a long transaction.
Even state-of-the-art GC does not sufficiently reduce the number of visible versions if there is only a single long transaction.
This is because modern protocols assume that transaction execution is one-shot, and that the transactions are relatively short (e.g., 16 operations for original YCSB).
Long transactions except read-only ones have been ignored in modern transaction studies.

To generate a long transaction, we added an artificial delay at the end of the read phase.
Both long and short transactions used the same number of operations with the same read-write ratio.
One worker thread executed the long transaction,
and the remaining worker threads executed the short transactions.
The skew was set to 0 so contention in record accesses rarely occurred and thus did not affect performance, even though there was a long transaction.

We measured the performance of Cicada under the workload described above, varying the {\sf RapidGC} interval settings and the delay added to the long transaction.
As shown in Fig.~\ref{fig:comp_cicada-ninline_ycsbA_tuple1m_gci1us-1s_tps},
performance saturated when a delay was inserted.
Saturation occurred when the GC interval was the same as the added delay.
For example, the light blue line includes a long transaction with a 1 ms delay, and performance saturated when the GC interval was 1 ms.
Similar behaviors were observed with longer delays.
This is because current GC methods do not collect visible versions that may be read by active long transactions.
The current GC scheme does not collect the versions until the transaction finishes.
We consider that this limitation is the primary hurdle to improving MVCC performance by reducing the number of visible versions.
We could not obtain results for a 1 s delay because such a delay requires a huge amount of memory, which causes the Linux OOM (out-of-memory) killer to kill the process.

\subsection{Aggressive Garbage Collection}
\label{sec:discussion-towards-version-lifetime-control}
From the limitation of the current GC scheme described above,
we suggest a novel GC scheme, {\sf AggressiveGC}, that aggressively collects versions beyond the current ones to deal with long transactions.
For example, the multi-version timestamp ordering (MVTO) protocol could be integrated with a GC method that aggressively collects visible versions.
It could make some versions non-visible even though active or future transactions need to read them.
Such a protocol might incur read operation failures unlike conventional MVTO, which could be handled by aborting the transaction and retrying it with a new timestamp.
Restricting the number of versions in kVSR~\cite{10.5555/645478.654954} and 2V2PL~\cite{10.1145/320141.320146,10.1145/582318.582330} has been discussed.
However, only the latest contiguous versions are kept there, so this approach is less flexible than our suggested scheme.
We claim that the visible versions do not need to be contiguous and the number of versions can be flexible depending on the context.
An interesting topic in our proposed scheme is the risk of starvation.
One way to mitigate this problem is to manage the priorities among transactions such as wait-die~\cite{10.1145/320251.320260}, which has not yet been discussed in modern MVCC studies.

We suggest two optimizations for write operations in terms of aggressive protocols.
The first is version overwriting, i.e., creating a new version by overwriting the memory segment of the previous version, which becomes non-visible at the same time, as is done in SVCC protocols.
Version overwriting is efficient because two operations are combined into one operation.
The second is non-visible write, i.e., making versions non-visible from the beginning of their life.
The idea of non-visible write was originally proposed as the Thomas write rule~\cite{10.1145/320071.320076} and recently generalized as the non-visible write rule (NWR)~\cite{nakazono2019invisiblewriterule} to deal with blind writes.
Novel version lifetime management is a promising way to improve MVCC protocols.

\underline{\bf Insight 6:}~
Even state-of-the-art GC cannot hold down the number of versions in MVCC protocols if a single long transaction is mixed into the workload.
An aggressive approach can solve this problem by aggressively changing the version state to non-visible for both reads and writes, even if transactions still require the state.

\ifvldb
\else
\\
\\
\fi
\section{Related Work}
\label{sec:Related Work}
Yu et al.~\cite{yu2014staring} evaluated CC protocols using DBx1000~\cite{dbx1000}, which is open source.
They evaluated the scalability of CC protocols using a real machine and an emulator.
Three recent protocols~
\cite{
  kim2016ermia,
  lim2017cicada,
  wang2016mostly}
~supported in CCBench, are not included in DBx1000~\cite{dbx1000}.
Wu et al.~\cite{wu2017empirical} empirically evaluated MVCC protocols using Peloton~\cite{peloton}, which is open source.
They evaluated not only scalability but also the contention effect, read ratio, attributes, memory allocation, and index.
They did not evaluate SVCC and the modern MVCC protocols evaluated in this paper~
\cite{
  lim2017cicada,
  tu2013speedy,
  wang2016mostly,
  yu2016tictoc}.
Appuswamy et al.~\cite{trireme} evaluated CC protocols in four types of architecture using Trireme, which is not open source.
They determined that the shared-everything architecture is still the best option for contention-tolerant in-memory transaction engines.
CC protocols for distributed systems have been evaluated elsewhere \cite{10.14778/3055540.3055548,wang2020comprehensive}; this paper focuses on a single many-core architecture.

%
%
%
Whereas previous studies mostly evaluated scalability and did not explore the behavior of protocols when thread parallelism was set to a high degree~
\cite{
  kim2016ermia,   
  lim2017cicada,  
  tu2013speedy,   
  wang2016mostly, 
  yu2016tictoc,   
  Repair,
  cavalia_paper,
  Healing,
  ACC_Tang,
  kimura2015foedus,
  IC3,
  STOv2,
  Strife,
  BCC,
  EWV,
  AOCC_Guo,
  tx_batch,
  SGT},
  we fixed the thread parallelism at 224 and analyzed protocols for various settings.
%
We classified a variety of methods on the basis of three performance factors: cache, delay, and version lifetime.
This analysis let us identify three new optimization methods.

\section{Conclusion}
\label{sec:Conclusion}
%
%
Using CCBench, we analyzed concurrency control protocols and optimization methods
for various settings of the workload parameters with the number of threads fixed at 224,
whereas previous studies mostly focused on thread scalability,
and none of them explored the space we analyzed.
%
%
%
We classified versatile optimization methods on the basis of three performance factors: cache, delay, and version lifetime.
Through the analysis of protocols with CCBench, we gained six insights.
{\bf I1:}~The performance of optimistic concurrency control for a read-only workload rapidly degrades as cardinality increases even without L3 cache misses.
{\bf I2:}~Silo outperforms TicToc for write-intensive workloads, which is attributed to {\sf InvisibleReads} for unskewed high cardinality cases.
{\bf I3:}~The effectiveness of two approaches to coping with conflict ({\sf Wait} and {\sf NoWait}) depends on the situation.
{\bf I4:}~Extra reads can regulate contention.
{\bf I5:}~Results produced from mixed implementations may be inconsistent with the theory.
{\bf I6:}~Even a state-of-the-art garbage collection method {\sf RapidGC } cannot improve the performance of multi-version concurrency control if there is a single long transaction mixed into the workload
On the basis of {\bf I4}, we defined the {\sf ReadPhaseExtension} optimization in which an artificial delay is added to the read phase. On the basis of {\bf I6}, we defined the {\sf AggressiveGC} optimization in which even visible versions are collected.

\ifvldb
\else
In future work, we will support TPC-C full-mix other than Silo, 
\fi
and to include logging and recovery modules based on our preliminary studies~\cite{tanabe2018cicada-pwal2,nakamuraTicToc}.
The code for CCBench and all the data in this paper are available online at GitHub~\cite{thawkccbench}.
We expect that CCBench will help to advance transaction processing research.


\newpage

\ifvldb

\fi
\bibliographystyle{abbrv}
\bibliography{ref}  

\begin{thebibliography}{10}

\bibitem{cc_format}
{CCBench Developer Guide}.
\newblock \url{https://github.com/thawk105/ccbench/tree/master/cc_format}.

\bibitem{thawkccdata}
{CCBench Experimental Data}.
\newblock \url{https://github.com/thawk105/ccdata}.

\bibitem{occ_nemoto}
{CCBench OCC}.
\newblock \url{https://github.com/thawk105/ccbench/tree/master/occ}.

\bibitem{cavalia}
{Code of Cavalia}.
\newblock \url{https://github.com/Cavalia}.

\bibitem{thawkccbench}
{Code of CCBench}.
\newblock \url{https://github.com/thawk105/ccbench}.

\bibitem{cicada_git}
{Code of Cicada}.
\newblock \url{https://github.com/efficient/cicada-engine}.

\bibitem{dbx1000}
{Code of DBx1000}.
\newblock \url{https://github.com/yxymit/DBx1000}.

\bibitem{dbx1000_cicada_git}
{Code of DBx1000 Extention for Cicada}.
\newblock \url{https://github.com/efficient/cicada-exp-sigmod2017-DBx1000}.

\bibitem{ermia_git}
{Code of ERMIA}.
\newblock \url{https://github.com/ermia-db/ermia}.

\bibitem{foedus_git}
{Code of FOEDUS}.
\newblock \url{https://github.com/hkimura/foedus_code}.

\bibitem{masstree_beta}
{Code of Masstree}.
\newblock \url{https://github.com/kohler/masstree-beta}.

\bibitem{peloton}
{Code of Peloton}.
\newblock \url{https://pelotondb.io}.

\bibitem{SGT_git}
{Code of SGT}.
\newblock \url{https://github.com/durner/No-False-Negatives}.

\bibitem{silo_git}
{Code of Silo}.
\newblock \url{https://github.com/stephentu/silo}.

\bibitem{sto_git}
{Code of STO}.
\newblock \url{https://readablesystems.github.io/sto}.

\bibitem{gflags}
{gflags}.
\newblock \url{https://github.com/gflags/gflags}.

\bibitem{nemotoccbench}
{How to Extend CCBench}.
\newblock \url{https://medium.com/@jnmt}.

\bibitem{bug_masstree_cast}
{Masstree Debug about Cast}.
\newblock
  \url{https://github.com/thawk105/masstree-beta/commit/d4bcf7711dc027818b1719a5a4c29aee547c58f6}.

\bibitem{bug_masstree_longkey_commit}
{Masstree Debug about string\_slice.hh}.
\newblock
  \url{https://github.com/thawk105/masstree-beta/commit/77ef355f868a6db4eac7b44669c508d8db053502}.

\bibitem{bug_masstree_longkey}
{Masstree when Controling Key Length is 9}.
\newblock \url{https://github.com/kohler/masstree-beta/issues/42}.

\bibitem{mimalloc}
{mimalloc}.
\newblock \url{https://github.com/microsoft/mimalloc}.

\bibitem{tpcc}
{The Transaction Processing Council. TPC-C Benchmark (Revision 5.11), February
  2011.}

\bibitem{tpcc_runner_git}
{tpcc-runner}.
\newblock \url{https://github.com/wattlebirdaz/tpcc-runner}.

\bibitem{trireme}
R.~Appuswamy, A.~G. Anadiotis, D.~Porobic, M.~Iman, and A.~Ailamaki.
\newblock {Analyzing the Impact of System Architecture on the Scalability of
  {OLTP} Engines for High-Contention Workloads}.
\newblock {\em PVLDB}, 11(2):121--134, 2017.

\bibitem{10.1145/3399666.3399910}
T.~Bang, N.~May, I.~Petrov, and C.~Binnig.
\newblock {The Tale of 1000 Cores: An Evaluation of Concurrency Control on
  Real(Ly) Large Multi-Socket Hardware}.
\newblock In {\em DaMoN}, 2020.

\bibitem{10.1145/320141.320146}
R.~Bayer, H.~Heller, and A.~Reiser.
\newblock {Parallelism and Recovery in Database Systems}.
\newblock {\em ACM TODS}, 5(2):139--156, 1980.

\bibitem{berenson1995critique}
H.~Berenson, P.~Bernstein, J.~Gray, J.~Melton, E.~O'Neil, and P.~O'Neil.
\newblock {A Critique of ANSI SQL Isolation Levels}.
\newblock In {\em SIGMOD Record}, volume~24, pages 1--10, 1995.

\bibitem{10.1145/356842.356846}
P.~A. Bernstein and N.~Goodman.
\newblock {Concurrency Control in Distributed Database Systems}.
\newblock {\em ACM Comput. Surv.}, 13(2):185--221, 1981.

\bibitem{bernstein1987concurrency}
P.~A. Bernstein, V.~Hadzilacos, and N.~Goodman.
\newblock {Concurrency control and recovery in database systems}.
\newblock 1987.

\bibitem{clements2013radixvm}
A.~T. Clements, M.~F. Kaashoek, and N.~Zeldovich.
\newblock {RadixVM: Scalable address spaces for multithreaded applications}.
\newblock In {\em EuroSys}, pages 211--224, 2013.

\bibitem{Repair}
M.~Dashti, S.~Basil~John, A.~Shaikhha, and C.~Koch.
\newblock {Transaction Repair for Multi-Version Concurrency Control}.
\newblock In {\em SIGMOD Conf.}, pages 235--250, 2017.

\bibitem{oltpbench}
D.~E. Difallah, A.~Pavlo, C.~Curino, and P.~Cudre-Mauroux.
\newblock {OLTP-Bench: An Extensible Testbed for Benchmarking Relational
  Databases}.
\newblock {\em PVLDB}, 7(4):277--288, 2013.

\bibitem{tx_batch}
B.~Ding, L.~Kot, and J.~Gehrke.
\newblock {Improving Optimistic Concurrency Control through Transaction
  Batching and Operation Reordering}.
\newblock {\em PVLDB}, 12(2):169--182, 2018.

\bibitem{SGT}
D.~Durner and T.~Neumann.
\newblock {No False Negatives: Accepting All Useful Schedules in a Fast
  Serializable Many-Core System}.
\newblock In {\em ICDE}, pages 734--745, 2019.

\bibitem{eswaran1976notions}
K.~P. Eswaran, J.~N. Gray, R.~A. Lorie, and I.~L. Traiger.
\newblock {The Notions of Consistency and Predicate Locks in a Database
  System}.
\newblock {\em Comm. ACM}, 19(11):624--633, 1976.

\bibitem{EWV}
J.~M. Faleiro, D.~J. Abadi, and J.~M. Hellerstein.
\newblock {High Performance Transactions via Early Write Visibility}.
\newblock {\em PVLDB}, 10(5):613--624, 2017.

\bibitem{gray1994quickly}
J.~Gray, P.~Sundaresan, S.~Englert, K.~Baclawski, and P.~J. Weinberger.
\newblock {Quickly generating billion-record synthetic databases}.
\newblock In {\em SIGMOD Record}, volume~23, pages 243--252, 1994.

\bibitem{AOCC_Guo}
J.~Guo, P.~Cai, J.~Wang, W.~Qian, and A.~Zhou.
\newblock {Adaptive Optimistic Concurrency Control for Heterogeneous
  Workloads}.
\newblock {\em PVLDB}, 12(5):584--596, 2019.

\bibitem{10.14778/3055540.3055548}
R.~Harding, D.~Van~Aken, A.~Pavlo, and M.~Stonebraker.
\newblock {An Evaluation of Distributed Concurrency Control}.
\newblock {\em PVLDB}, 10(5):553--564, 2017.

\bibitem{STOv2}
Y.~Huang, W.~Qian, E.~Kohler, B.~Liskov, and L.~Shrira.
\newblock {Opportunities for Optimism in Contended Main-Memory Multicore
  Transactions}.
\newblock {\em PVLDB}, 13(5):629--642, 2020.

\bibitem{aether}
R.~Johnson, I.~Pandis, R.~Stoica, M.~Athanassoulis, and A.~Ailamaki.
\newblock {Aether: a Scalable Approach to Logging}.
\newblock {\em PVLDB}, 3(1-2):681--692, 2010.

\bibitem{jung2013performance}
H.~Jung, H.~Han, A.~Fekete, U.~R{\"o}hm, and H.~Y. Yeom.
\newblock {Performance of Serializable Snapshot Isolation on Multicore
  Servers}.
\newblock In {\em DASFAA}, pages 416--430, 2013.

\bibitem{CompileVector}
T.~Kersten, V.~Leis, A.~Kemper, T.~Neumann, A.~Pavlo, and P.~Boncz.
\newblock {Everything You Always Wanted to Know about Compiled and Vectorized
  Queries but Were Afraid to Ask}.
\newblock {\em PVLDB}, 11(13):2209--2222, 2018.

\bibitem{kim2016ermia}
K.~Kim, T.~Wang, R.~Johnson, and I.~Pandis.
\newblock {ERMIA: Fast Memory-Optimized Database System for Heterogeneous
  Workloads}.
\newblock In {\em SIGMOD Conf.}, pages 1675--1687, 2016.

\bibitem{kimura2015foedus}
H.~Kimura.
\newblock {FOEDUS: OLTP engine for a thousand cores and NVRAM}.
\newblock In {\em SIGMOD Conf.}, pages 691--706, 2015.

\bibitem{10.14778/2095686.2095689}
P.~Larson, S.~Blanas, C.~Diaconu, C.~Freedman, J.~M. Patel, and M.~Zwilling.
\newblock {High-Performance Concurrency Control Mechanisms for Main-Memory
  Databases}.
\newblock {\em PVLDB}, 5(4):298--309, 2011.

\bibitem{lim2017cicada}
H.~Lim, M.~Kaminsky, and D.~G. Andersen.
\newblock {Cicada: Dependably fast multi-core in-memory transactions}.
\newblock In {\em SIGMOD Conf.}, pages 21--35, 2017.

\bibitem{mao2012cache}
Y.~Mao, E.~Kohler, and R.~T. Morris.
\newblock {Cache Craftiness for Fast Multicore Key-Value Storage}.
\newblock In {\em EuroSys}, pages 183--196, 2012.

\bibitem{marathe2004design}
V.~J. Marathe, W.~N. Scherer, and M.~L. Scott.
\newblock {Design Tradeoffs in Modern Software Transactional Memory Systems}.
\newblock In {\em LCR}, pages 1--7, 2004.

\bibitem{mellor1991scalable}
J.~M. Mellor-Crummey and M.~L. Scott.
\newblock {Scalable Reader-Writer Synchronization for Shared-Memory
  Multiprocessors}.
\newblock In {\em SIGPLAN Notices}, volume~26, pages 106--113, 1991.

\bibitem{10.5555/645478.654954}
T.~Morzy.
\newblock {The Correctness of Concurrency Control for Multiversion Database
  Systems with Limited Number of Versions}.
\newblock In {\em ICDE}, pages 595--604, 1993.

\bibitem{nakamuraTicToc}
Y.~Nakamura, H.~Kawashima, and O.~Tatebe.
\newblock {Integration of TicToc Concurrency Control Protocol with Parallel
  Write Ahead Logging Protocol}.
\newblock {\em Journal of Network Computing}, 9(2):339--353, 2019.

\bibitem{nakazono2019invisiblewriterule}
S.~Nakazono, H.~Uchiyama, Y.~Fujiwara, Y.~Nakamura, and H.~Kawashima.
\newblock {NWR: Rethinking Thomas Write Rule for Omittable Write Operations}.
\newblock \url{http://arxiv.org/abs/1904.08119}, 2020.

\bibitem{Doppel}
N.~Narula, C.~Cutler, E.~Kohler, and R.~Morris.
\newblock {Phase Reconciliation for Contended In-Memory Transactions}.
\newblock In {\em OSDI}, pages 511--524, 2014.

\bibitem{10.1145/2723372.2749436}
T.~Neumann, T.~M\"{u}hlbauer, and A.~Kemper.
\newblock {Fast Serializable Multi-Version Concurrency Control for Main-Memory
  Database Systems}.
\newblock In {\em SIGMOD Conf.}, pages 677--689, 2015.

\bibitem{PelotonConcept}
A.~Pavlo, G.~Angulo, J.~Arulraj, H.~Lin, J.~Lin, L.~Ma, P.~Menon, T.~C. Mowry,
  M.~Perron, I.~Quah, S.~Santurkar, A.~Tomasic, S.~Toor, D.~V. Aken, Z.~Wang,
  Y.~Wu, R.~Xian, and T.~Zhang.
\newblock {Self-Driving Database Management Systems}.
\newblock In {\em CIDR}, 2017.

\bibitem{Strife}
G.~Prasaad, A.~Cheung, and D.~Suciu.
\newblock {Handling Highly Contended {OLTP} Workloads Using Fast Dynamic
  Partitioning}.
\newblock In {\em SIGMOD Conf.}, pages 527--542, 2020.

\bibitem{10.1145/320251.320260}
D.~J. Rosenkrantz, R.~E. Stearns, and P.~M. Lewis.
\newblock {System Level Concurrency Control for Distributed Database Systems}.
\newblock {\em ACM TODS}, 3(2):178--198, 1978.

\bibitem{scott2001scalable}
M.~L. Scott and W.~N. Scherer.
\newblock {Scalable Queue-based Spin Locks with Timeout}.
\newblock In {\em SIGPLAN Notices}, volume~36, pages 44--52, 2001.

\bibitem{abortML}
Y.~Sheng, A.~Tomasic, T.~Zhang, and A.~Pavlo.
\newblock {Scheduling OLTP Transactions via Learned Abort Prediction}.
\newblock In {\em aiDM}, 2019.

\bibitem{10.1145/582318.582330}
R.~E. Stearns and D.~J. Rosenkrantz.
\newblock {Distributed Database Concurrency Controls Using Before-Values}.
\newblock In {\em SIGMOD Conf.}, pages 74--83, 1981.

\bibitem{tanabe2018cicada-pwal2}
T.~Tanabe, H.~Kawashima, and O.~Tatebe.
\newblock {Integration of Parallel Write Ahead Logging and Cicada Concurrency
  Control Method}.
\newblock In {\em BITS}, pages 291--296, 2018.

\bibitem{CormCC}
D.~Tang and A.~J. Elmore.
\newblock Toward coordination-free and reconfigurable mixed concurrency
  control.
\newblock In {\em ATC}, pages 809--822, 2018.

\bibitem{ACC_Tang}
D.~Tang, H.~Jiang, and A.~J. Elmore.
\newblock {Adaptive Concurrency Control: Despite the Looking Glass, One
  Concurrency Control Does Not Fit All}.
\newblock In {\em CIDR}, 2017.

\bibitem{10.1145/320071.320076}
R.~H. Thomas.
\newblock {A Majority Consensus Approach to Concurrency Control for Multiple
  Copy Databases}.
\newblock {\em ACM TODS}, 4(2):180--209, 1979.

\bibitem{tu2013speedy}
S.~Tu, W.~Zheng, E.~Kohler, B.~Liskov, and S.~Madden.
\newblock {Speedy Transactions in Multicore in-Memory Databases}.
\newblock In {\em SOSP}, pages 18--32, 2013.

\bibitem{wang2020comprehensive}
C.~Wang, K.~Huang, and X.~Qian.
\newblock {A Comprehensive Evaluation of RDMA-enabled Concurrency Control
  Protocols}.
\newblock \url{http://arxiv.org/abs/2002.12664}, 2020.

\bibitem{wang2015serial}
T.~Wang, R.~Johnson, A.~Fekete, and I.~Pandis.
\newblock {The Serial Safety Net: Efficient Concurrency Control on Modern
  Hardware}.
\newblock In {\em DaMoN}, 2015.

\bibitem{wang2017efficiently}
T.~Wang, R.~Johnson, A.~Fekete, and I.~Pandis.
\newblock {Efficiently Making (Almost) Any Concurrency Control Mechanism
  Serializable}.
\newblock {\em VLDB Journal}, 26(4):537--562, 2017.

\bibitem{wang2016mostly}
T.~Wang and H.~Kimura.
\newblock {Mostly-Optimistic Concurrency Control for Highly Contended Dynamic
  Workloads on a Thousand Cores}.
\newblock {\em PVLDB}, 10(2):49--60, 2016.

\bibitem{IC3}
Z.~Wang, S.~Mu, Y.~Cui, H.~Yi, H.~Chen, and J.~Li.
\newblock {Scaling Multicore Databases via Constrained Parallel Execution}.
\newblock In {\em SIGMOD Conf.}, pages 1643--1658, 2016.

\bibitem{weikum2001transactional}
G.~Weikum and G.~Vossen.
\newblock {\em {Transactional Information Systems}}.
\newblock Elsevier, 2001.

\bibitem{wu2017empirical}
Y.~Wu, J.~Arulraj, J.~Lin, R.~Xian, and A.~Pavlo.
\newblock {An empirical evaluation of in-memory multi-version concurrency
  control}.
\newblock {\em PVLDB}, 10(7):781--792, 2017.

\bibitem{Healing}
Y.~Wu, C.-Y. Chan, and K.-L. Tan.
\newblock {Transaction Healing: Scaling Optimistic Concurrency Control on
  Multicores}.
\newblock In {\em SIGMOD Conf.}, pages 1689--1704, 2016.

\bibitem{cavalia_paper}
Y.~Wu and K.-L. Tan.
\newblock {Scalable In-Memory Transaction Processing with HTM}.
\newblock In {\em ATC}, pages 365--377, 2016.

\bibitem{yu2014staring}
X.~Yu, G.~Bezerra, A.~Pavlo, S.~Devadas, and M.~Stonebraker.
\newblock {Staring into the Abyss: An Evaluation of Concurrency Control with
  One Thousand Cores}.
\newblock {\em PVLDB}, 8(3):209--220, 2014.

\bibitem{yu2016tictoc}
X.~Yu, A.~Pavlo, D.~Sanchez, and S.~Devadas.
\newblock {Tictoc: Time Traveling Optimistic Concurrency Control}.
\newblock In {\em SIGMOD Conf.}, pages 1629--1642, 2016.

\bibitem{BCC}
Y.~Yuan, K.~Wang, R.~Lee, X.~Ding, J.~Xing, S.~Blanas, and X.~Zhang.
\newblock {BCC: Reducing False Aborts in Optimistic Concurrency Control with
  Low Cost for in-Memory Databases}.
\newblock {\em PVLDB}, 9(6):504--515, 2016.

\end{thebibliography}
\newpage
\ifvldb
\else

\begin{appendix}

%
%
 %
%
%
%
\begin{figure*}[tb]
 \begin{tabular}{ccc}
  \begin{minipage}{0.33\hsize}
   \subcaptionbox{Warehouse count = 1.
   \label{fig:tpcc warehouse=1}
   }
   {\includegraphics[scale=0.55]{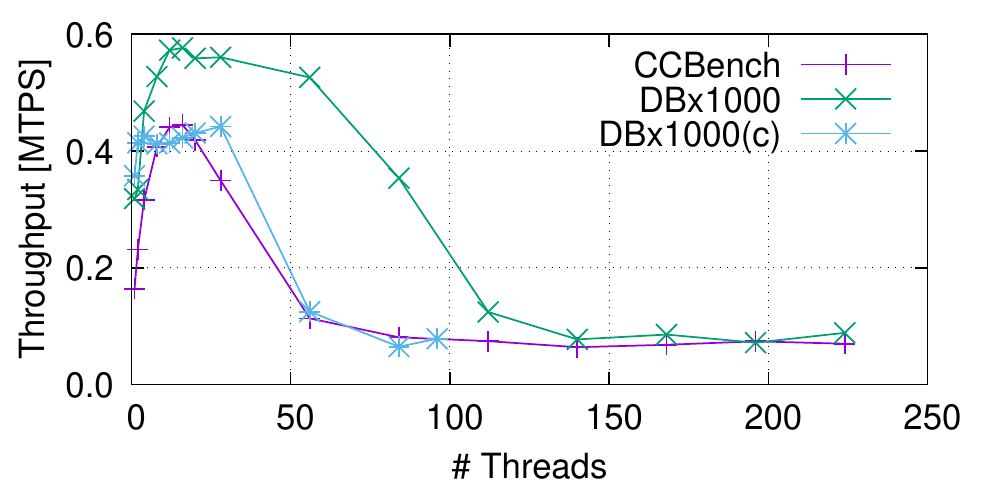}}
  \end{minipage}
  &
  \begin{minipage}{0.33\hsize}
   \subcaptionbox{Warehouses count = 4.
   \label{fig:tpcc warehouse=4}
   }
   {\includegraphics[scale=0.55]{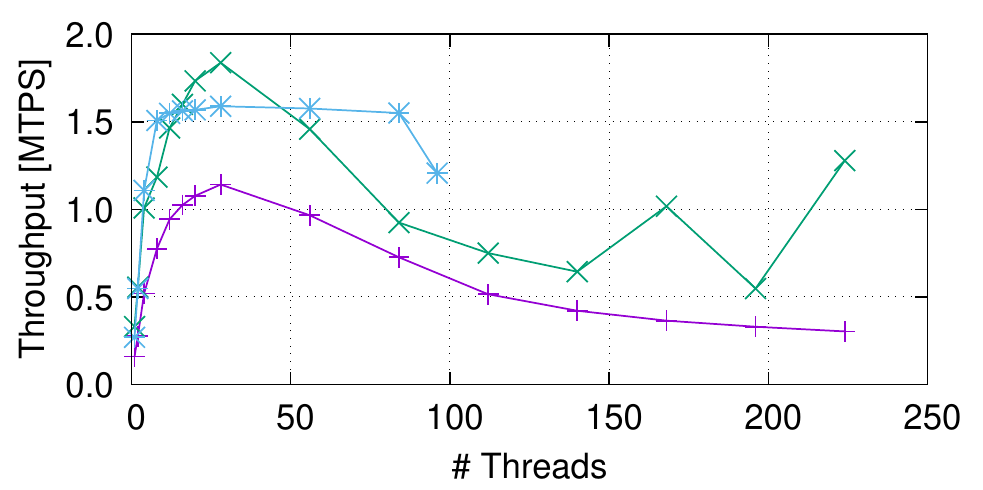}}
  \end{minipage}
  &
  \begin{minipage}{0.33\hsize}
   \subcaptionbox{Warehouse count = thread count.
   \label{fig:tpcc warehouse=thread}
   }
   {\includegraphics[scale=0.55]{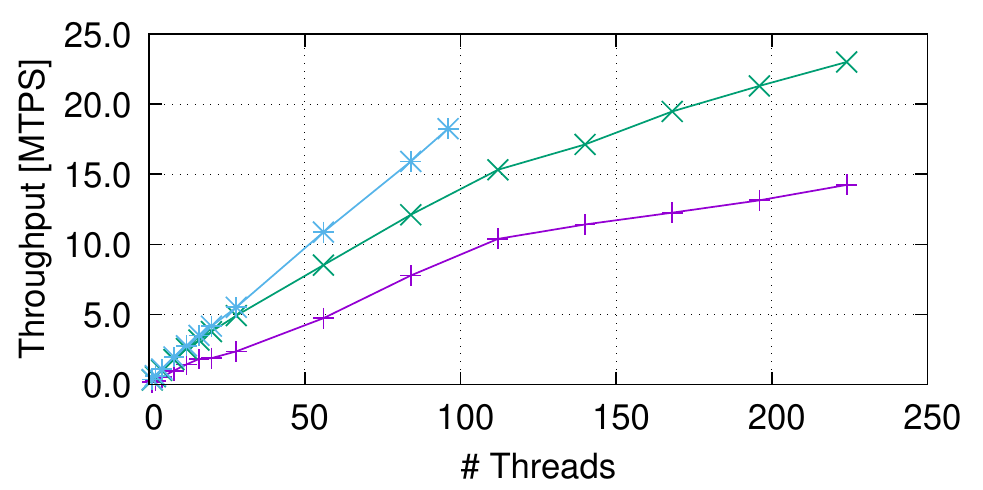}}
  \end{minipage}
 \end{tabular}
 \caption{TPC-C-NP-NoInsert (full scale).}
 \label{fig:TPC-C-NP-NoInsert (Full Scale).}
\end{figure*}

\begin{figure*}[tb]
 \begin{tabular}{ccc}
  \begin{minipage}{0.33\hsize}
   \subcaptionbox{Warehouse count = 1.
   \label{fig:tpcc warehouse=1(small)}
   }
   {\includegraphics[scale=0.55]{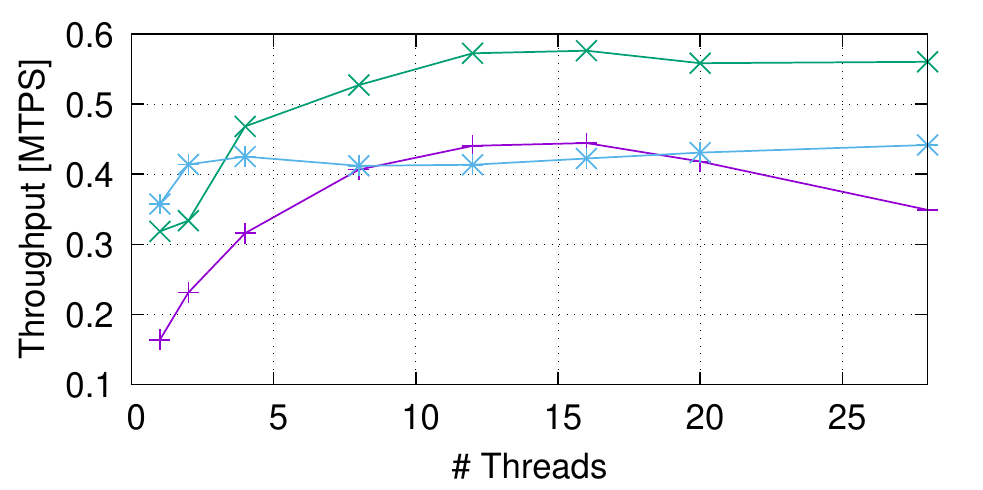}}
  \end{minipage}
  &
  \begin{minipage}{0.33\hsize}
     \subcaptionbox{Warehouse count = 4.
   \label{fig:tpcc warehouse=4(small)}
   }
     {\includegraphics[scale=0.55]{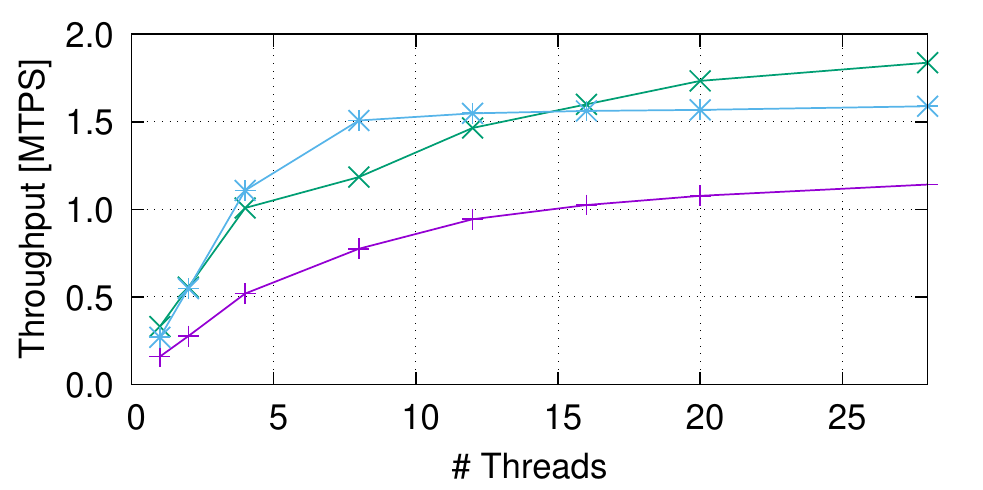}}
  \end{minipage}
  &
  \begin{minipage}{0.33\hsize}
   \subcaptionbox{Warehouse count = thread count.
   \label{fig:tpcc warehouse=thread(small)}
   }
   {\includegraphics[scale=0.55]{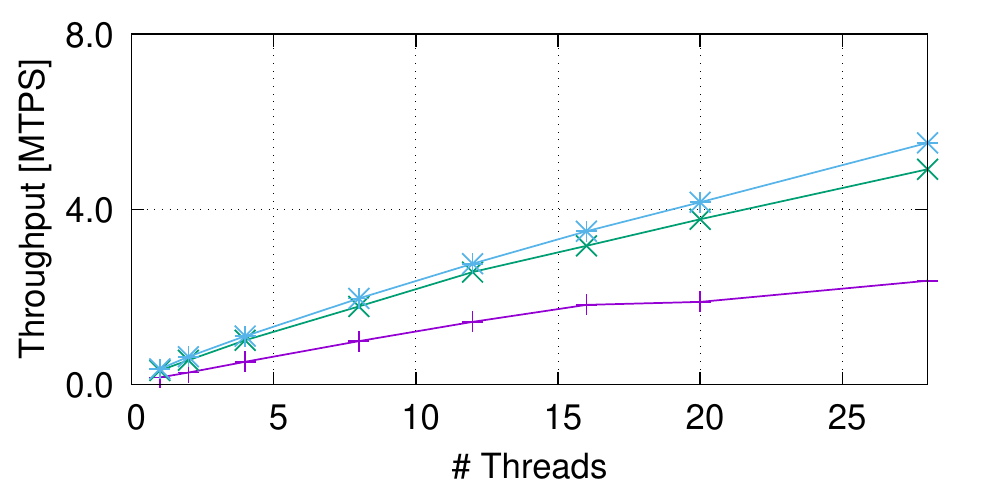}}
  \end{minipage}
 \end{tabular}
 \caption{TPC-C-NP-NoInsert (small scale).}
 \label{fig:TPC-C-NP-NoInsert (Small Scale).}
\end{figure*}

%
%
%
%
\begin{figure*}[tb]
 \begin{tabular}{ccc}
  \begin{minipage}{0.33\hsize}
   \subcaptionbox{Warehouse count = 1.
   \label{fig:tpcc warehouse=1, INSERT}
   }
   {\includegraphics[scale=0.55]{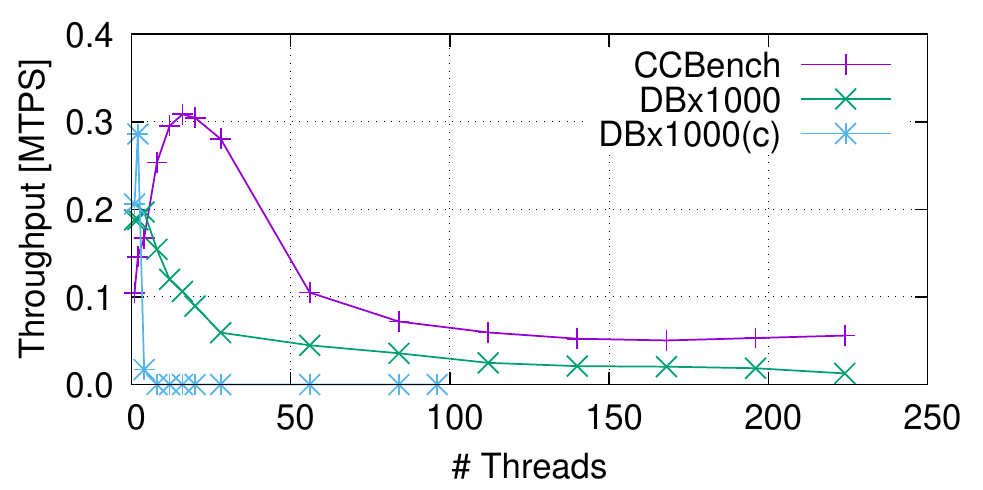}}
  \end{minipage}
  &
  \begin{minipage}{0.33\hsize}
   \subcaptionbox{Warehouses count = 4.
   \label{fig:tpcc warehouse=4, INSERT}
   }
   {\includegraphics[scale=0.55]{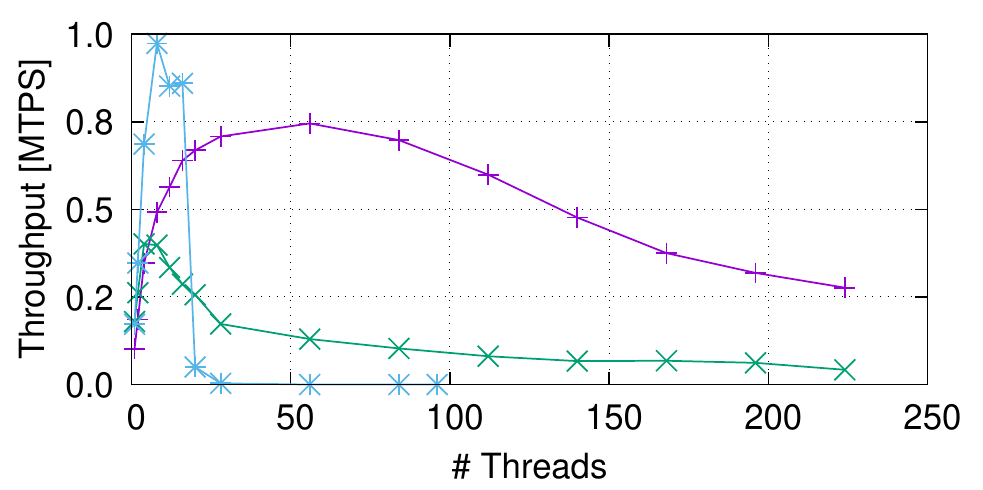}}
  \end{minipage}
  &
  \begin{minipage}{0.33\hsize}
   \subcaptionbox{Warehouse count = thread count.
   \label{fig:tpcc warehouse=thread, INSERT}
   }
   {\includegraphics[scale=0.55]{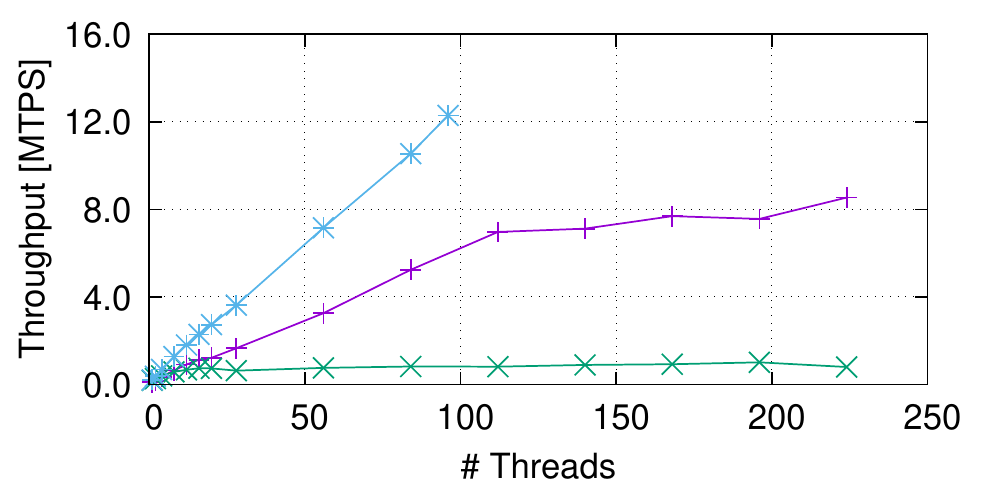}}
  \end{minipage}
 \end{tabular}
 \caption{TPC-C-NP-Insert (full scale).}
 \label{fig:TPC-C-NP-Insert (Full Scale).}
\end{figure*}

\begin{figure*}[tb]
 \begin{tabular}{ccc}
  \begin{minipage}{0.33\hsize}
   \subcaptionbox{Warehouse count = 1.
   \label{fig:tpcc warehouse=1(small), INSERT}
   }
   {\includegraphics[scale=0.55]{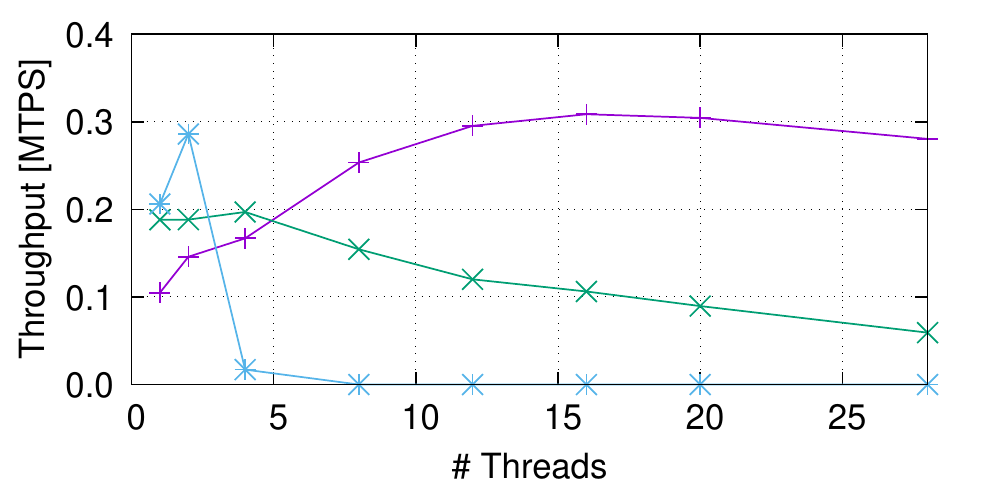}}
  \end{minipage}
  &
  \begin{minipage}{0.33\hsize}
   \subcaptionbox{Warehouse count = 4.
   \label{fig:tpcc warehouse=4(small), INSERT}
   }
   {\includegraphics[scale=0.55]{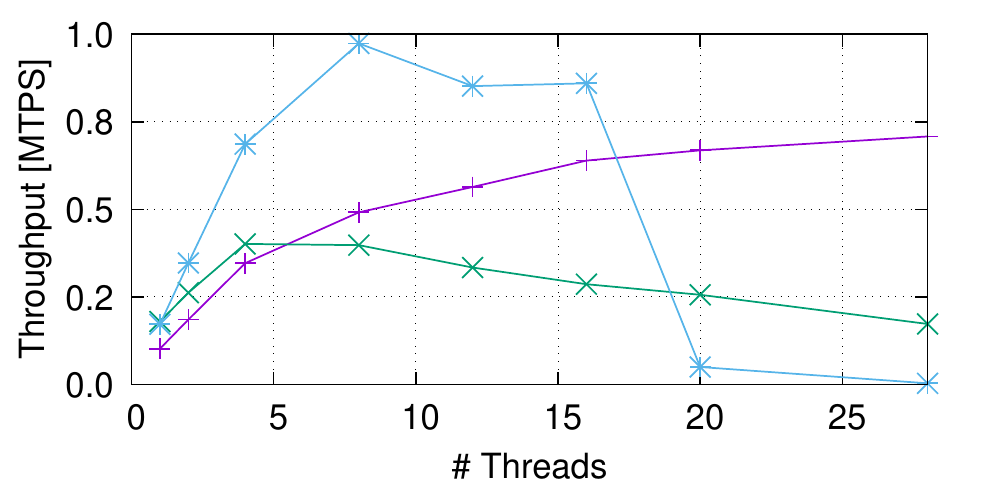}}
  \end{minipage}
  &
  \begin{minipage}{0.33\hsize}
   \subcaptionbox{Warehouse count = thread count.
   \label{fig:tpcc warehouse=thread(small), INSERT}
   }
   {\includegraphics[scale=0.55]{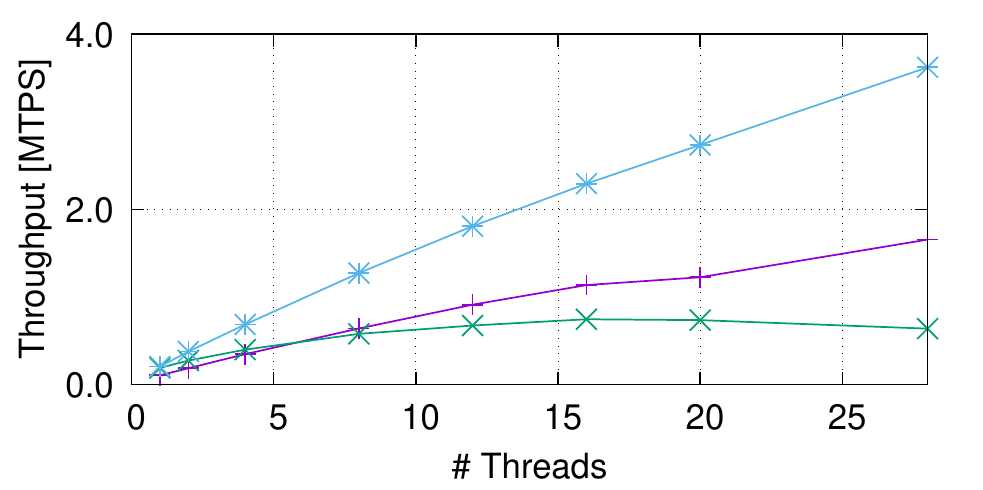}}
  \end{minipage}
 \end{tabular}
 \caption{TPC-C-NP-Insert (small scale).}
 \label{fig:TPC-C-NP-Insert (Small Scale).}
\end{figure*}

%
%
\begin{figure}[tb]
  \centering
  \includegraphics[scale=0.6]{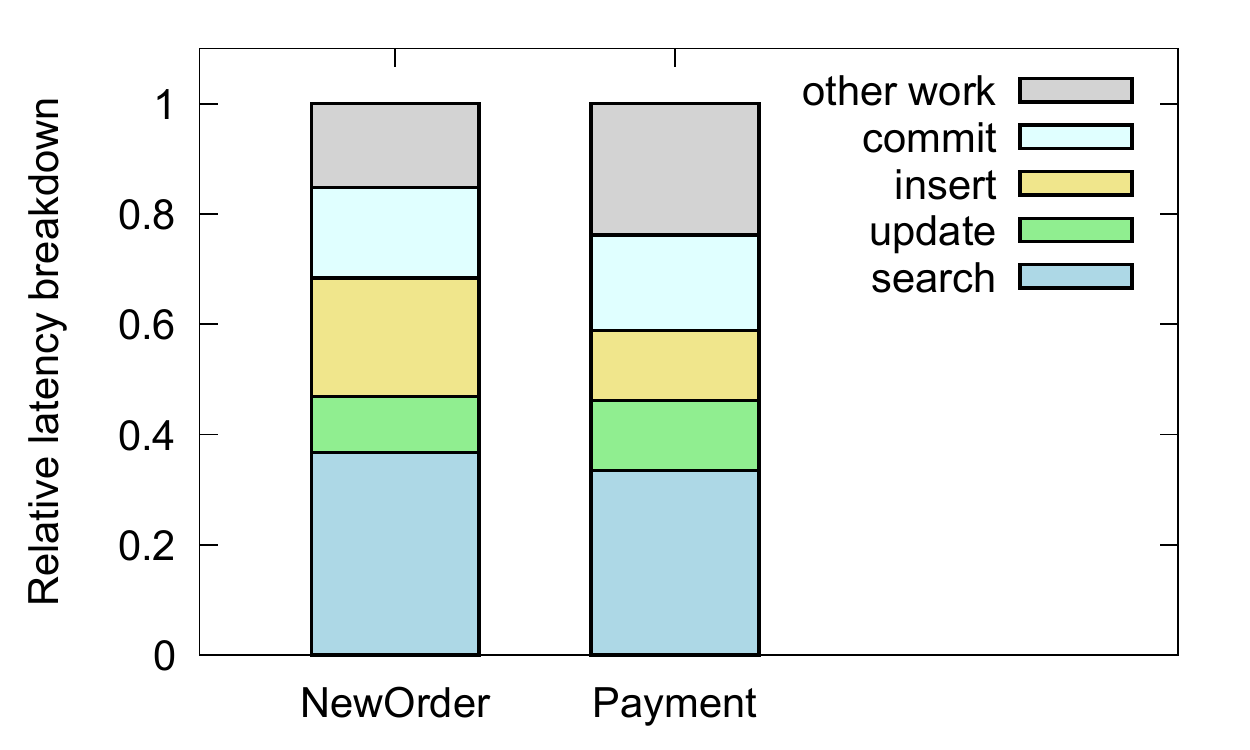}
 \caption{Breakdown of TPC-C-NP-Insert, 224 threads, 224 warehouses, and NewOrder and Payment transactions.}
 \label{fig:breakdown}
\end{figure}

%
%
\begin{figure}[tb]
 \centering
 \includegraphics[scale=0.6]{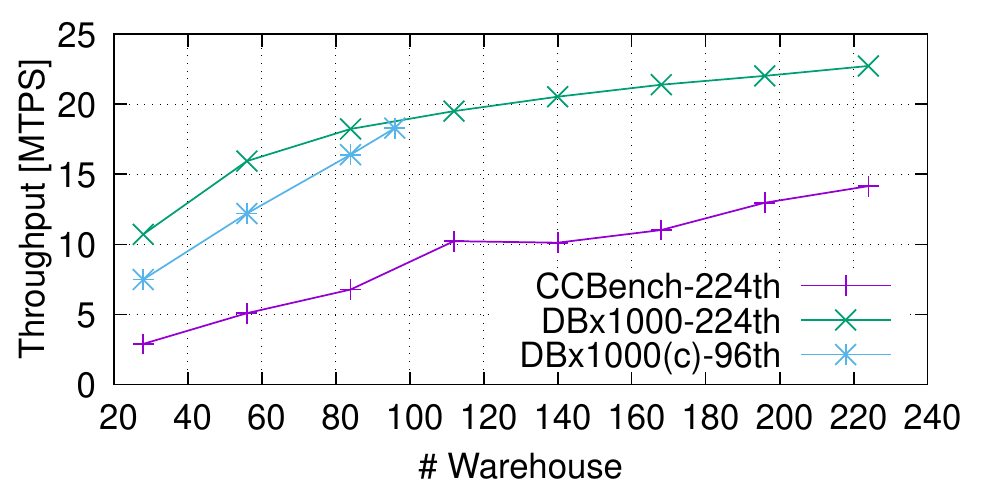}
 \caption{Varying warehouse count, NP-NoInsert.}
 \label{fig:Varying warehouse count, NP-NoInsert}
 \includegraphics[scale=0.6]{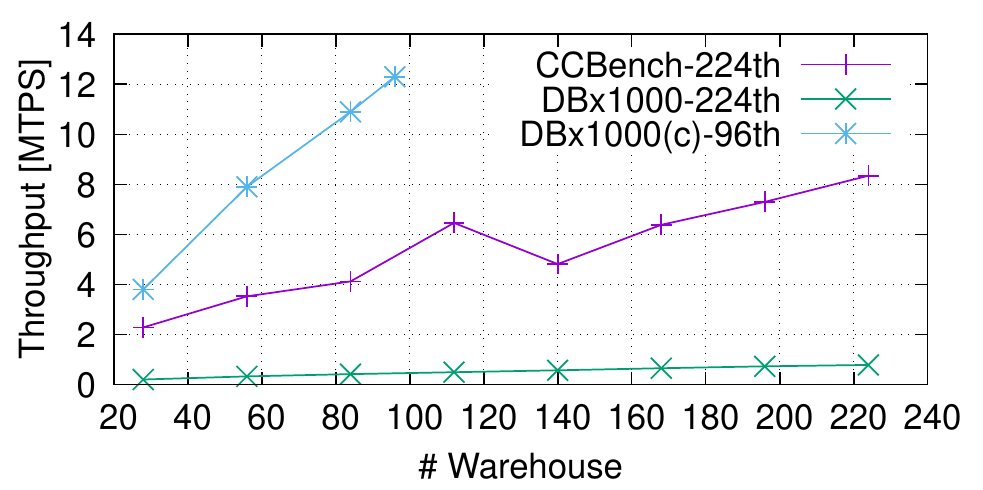}
 \caption{Varying warehouse count, NP-Insert.}
 \label{fig:Varying warehouse count, NP-Insert}
\end{figure}

 \begin{table}[tb]
 \begin{center}
  \caption{Supported TPC-C transactions and execution platforms in modern CC studies for a single node.
   {\sf Type} column indicates the type of research. Papers with {\sf Exp.} describe analytical studies, and those with {\sf Pro.} propose new protocols or optimization methods. 
  The {\sf TPC-C} column indicates the type of TPC-C transactions supported in the paper.
  {\sf NP} indicates {\sf NewOrder} and {\sf Payment}.
  {\sf Full} indicates the full-mix, and 
  $\alpha$ includes an original {\sf StockScan} transaction in addition to the full-mix.
  $\beta$ includes an original {\sf Reward} transaction in addition to the NP.
  $\phi$ indicates no TPC-C evaluation.
  The {\sf System} column shows the evaluation system used in the paper.
  No citation indicates that the code seems to be publicly unavailable.}
  \label{tab:Supported TPC-C}
\begin{tabular}[tb]{l|c|c|cc}
 {\rm Paper}   & {\rm Type} & {\rm TPC-C} & {\rm System} \\
 \hline
  \hline 
 MVCC Eval.~\cite{wu2017empirical} ($\alpha$)&\multirow{5}{*}{Exp.} & Full& {\tt Peloton}~\cite{peloton}\\
 \cline{3-4}
 CCBench                 &&\multirow{3}{*}{NP}& {\tt CCBench}~\cite{thawkccbench}\\
 Abyss~\cite{yu2014staring}                 & & & {\tt DBx1000}~\cite{dbx1000}\\
 1000 Cores~\cite{10.1145/3399666.3399910}  & & & {\tt DBx1000}~\cite{dbx1000}\\
 \cline{3-4}
 Trireme~\cite{trireme}                     &     & $\phi$ & Original\\
 \cline{2-4}
 Repair~\cite{Repair} &\multirow{20}{*}{Pro.}&\multirow{12}{*}{Full}& {\tt Cavalia}~\cite{cavalia}\\
 Healing~\cite{Healing}                   &  &  & {\tt Cavalia}~\cite{cavalia}\\
 HTCC~\cite{cavalia_paper}                &  &  & {\tt Cavalia}~\cite{cavalia}\\
 Cicada~\cite{lim2017cicada}              &  &  & {\tt Cicada}~\cite{cicada_git}\\
 ACC~\cite{ACC_Tang}                      &  &  &{\tt Doppel}~\cite{Doppel}\\
 Latch-free SSN~\cite{wang2017efficiently}&  &  &{\tt ERMIA}~\cite{ermia_git}\\
 FOEDUS~\cite{kimura2015foedus}           &  &  & {\tt FOEDUS}~\cite{foedus_git}\\
 MOCC~\cite{wang2016mostly}               &  &  & {\tt FOEDUS}~\cite{foedus_git}\\
 Silo~\cite{tu2013speedy}                 &  &  & {\tt Silo}~\cite{silo_git}\\
 IC3~\cite{IC3}                           &  &  & {\tt Silo}~\cite{silo_git}\\
 STOv2~\cite{STOv2}                       &  &  & {\tt STO}~\cite{sto_git}\\
 CormCC~\cite{CormCC}                     &  &  & Original\\
 \cline{3-4}
 TicToc~\cite{yu2016tictoc} &&\multirow{5}{*}{NP}& {\tt DBx1000}~\cite{dbx1000}\\
 Strife~\cite{Strife}                     &    &   & {\tt DBx1000}~\cite{dbx1000}\\
 AOCC~\cite{AOCC_Guo}($\beta$)            &    &   & {\tt DBx1000}~\cite{dbx1000}\\
 EWV~\cite{EWV}                           &    &   & Original\\
 BCC~\cite{BCC}                           &    &   & {\tt Silo}~\cite{silo_git}\\
 \cline{3-4}
 Batching~\cite{tx_batch}       &  &\multirow{3}{*}{$\phi$}& {\tt Cicada}~\cite{cicada_git}\\
 SGT~\cite{SGT}                 &  &  & Original~\cite{SGT_git}\\
 Doppel~\cite{Doppel}           &  &  & Original\\
\end{tabular}
 \end{center}
 \end{table}

 \section{Evaluation of a Subset of TPC-C}
 \label{sec:Evaluation of a Subset of TPC-C}
 The TPC-C benchmark emulates the workload of a wholesale supplier and is the current industry standard for evaluating transaction processing systems~\cite{tpcc}.
 Here, we introduce how previous studies and CCBench support TPC-C;
 describe the results of its evaluation, including a comparison with other systems~\cite{dbx1000,dbx1000_cicada_git};
 analyze the effect of index structures on the performance.

 %

\subsection{TPC-C-NP}
  \label{sec:Supported Transactions}
 TPC-C or its variants have been widely used to evaluate concurrency control protocols.
 Some studies fully support all five transactions, whereas some support only a subset of them.
 We summarize the studies in Table \ref{tab:Supported TPC-C}.
 Most of the papers with new proposals evaluated the full-mix.
 There were three experimental papers, among which only Wu et al.~\cite{wu2017empirical} evaluated the full-mix using Peloton~\cite{peloton}.
 Appuswamy et al.~\cite{trireme} did not evaluate TPC-C.
 Finally, Yu et al.~\cite{yu2014staring} and Bang et al.~\cite{10.1145/3399666.3399910} evaluated only {\tt New-Order} and {\tt Payment} from among the five transactions using DBx1000~\cite{dbx1000}.

 %
 %
 %
 Among the CC protocols, CCBench currently supports only Silo.
 %
 Here, we evaluate {\tt New-Order} and {\tt Payment} transactions in CCBench and validate the correctness of its implementation by comparing it with other platforms~\cite{dbx1000,dbx1000_cicada_git}.
 %

\color{black}
 %
 %
 In this Appendix, we denote the subset of the TPC-C workload that includes only {\tt New-Order} and {\tt Payment} transactions as {\sf NP}
 and illustrate an evaluation of two variants of {\sf NP}, denoted by {\sf NP-NoInsert} and {\sf NP-Insert}.
 They were supported by DBx1000, and both were evaluated by Bang et al.~\cite{10.1145/3399666.3399910}.

 {\sf NP-NoInsert} does not include any insertion operations.
 NP originally includes insertions into the {\tt Order, Order-Line, New-Order} tables in the {\tt New-Order} transaction and insertions into the {\tt History} table in the {\tt Payment} transaction.
 However, from the viewpoint of semantics, it is possible to omit these insertions.
 Records inserted into the {\tt History} table by the {\tt Payment} transaction are not handled thereafter (i.e., other transactions do not perform CRUD operations on the {\tt History} table). Records inserted into the {\tt Orders}, {\tt New-Order}, {\tt Order-Line} tables by {\tt New-Order} transaction are not read, updated, or deleted by the {\tt Payment} transaction. Therefore, {\sf NP} can omit all insertion operations.

 {\sf NP-Insert} includes insertion operations that are omitted in {\sf NP-NoInsert}.
 To run {\sf NP-NoInsert} or {\sf NP-Insert}, different functions should be implemented.
 We illustrate such differences in Table~\ref{tpcc_func}.
 As shown in the table, the required functions for {\sf NP-NoInsert} are the same as those for a part of YCSB.
 To implement the {\sf NP-Insert}, an insertion operation is necessary.
 \subsection{Implementation and Settings}
 \label{sec:Impl}
 %
 %
 We implemented TPC-C client codes with reference to the original {\bf DBx1000} system~\cite{dbx1000} and an extension of DBx1000 by the Cicada team~\cite{dbx1000_cicada_git}, which is denoted by {\bf \seqsplit{DBx1000(C)}}.
 %
 We implemented the {\sf NP} in CCBench as follows.
 First,
 %
 %
 for the access method, we used a Masstree implementation that is available online~\cite{masstree_beta} with some modifications (fixing bugs regarding the cast~\cite{bug_masstree_cast} and a long key with more than 9 bytes~\cite{bug_masstree_longkey,bug_masstree_longkey_commit}).
 %
 All nine tables were searchable with the primary key.
 The {\tt History} table did not originally need primary keys,
 but the table was stored using Masstree.
 Thus, it used dummy primary keys using the scalable, unique identifier generator that we developed.
The primary keys for all tables were encoded into 8 bytes in CCBench.
%

{\sf NP} requires a secondary index on the {\tt Customer} table with {\tt c\_w\_id}, {\tt c\_d\_id}, and {\tt c\_last} columns.
We stored multiple customer primary keys in a {\tt std::vector} container for each secondary key.
 %
 The size of the secondary keys was at most 20 bytes, 2 bytes for {\tt c\_w\_id}, 1 byte for {\tt c\_d\_id}, and up to 17 bytes for {\tt c\_last}.
Our {\sf NP} implementation is publicly available on GitHub~\cite{thawkccbench}.

 The settings of the other platforms were as follows.
 DBx1000 separates
 the table data structures from the primary index structures and can omit primary indexes for tables if unnecessary.
 We did not use the B+tree index owing to its low performance; therefore, we used a hash index for all indexes in our experiments.
 DBx1000 was configured by omitting an unnecessary primary index in {\sf NP}.
 {\tt Order}, {\tt New-Order}, {\tt Order-Line}, and {\tt History} tables had no primary index.
 The {\tt Customer} secondary index composed of the {\tt (c\_w\_id, c\_d\_id, c\_last)} key was encoded into 8 bytes.
 A pointer to the records was retrieved using the encoded key through the hash index, and was accessed through the linked list.
 \seqsplit{DBx1000(C)} also separates the tables and indexes.
 It properly uses the Masstree and hash index.

%
%
 \begin{table}[tb]
 \begin{center}
  \caption{Workload and required functions.
  YCSB' indicates only A, B, C, and F, which are key-wise.
  It does not include insert (required by D) or range query (required by E).
  Further, NP-NoInsert does not include any insertion operations.
  NP originally includes insertions to
  {\tt Order, Order-Line, and New-Order}
  tables in {\tt New-Order} transaction
  and insertions to {\tt History} table in Payment transaction.
  %
  NP-Insert includes insertion operations omitted in NP-NoInsert.
  Phantom avd. indicates phantom avoidance.
  Each cell denotes whether the corresponding function is required.
  }
  \label{tpcc_func}
  \begin{small}
  \begin{tabular}[tb]{l|cccc}
    & YCSB' & NP-NoInsert & NP-Insert & Full-mix\\
   \hline
   Insertion    & No & No & Yes & Yes\\
   Deletion     & No & No & No  & Yes\\
   Range search & No & No & No  & Yes\\
   Phantom avd. & No & No & No  & Yes\\
  \end{tabular}
  \end{small}
 \end{center}
 \end{table}

\subsection{Analysis}
\label{sec:Result}
We used the experimental environment described in $\S$\ref{sec:Environment}.
Each worker thread chose either a {\tt New-Order} or {\tt Payment} transaction at random in a 50:50 ratio each time.

\subsubsection{Varying Thread Count}
\label{sec:Varying Thread Count}
We evaluated the workloads by varying the thread count.

 %
 %
 {\bf NP-NoInsert: }
 For \seqsplit{DBx1000(C)}, we set a value of {\tt false} to two parameters,
 {\tt TPCC\_INSERT\_ROWS} and {\tt TPCC\_INSERT\_INDEX}.
 We commented out the code of insertion operations for DBx1000.
 We set the corresponding command-line argument for CCBench to omit insertion operations.
 The results are shown in Figs. \ref{fig:TPC-C-NP-NoInsert (Full Scale).} and \ref{fig:TPC-C-NP-NoInsert (Small Scale).}.
 Fig. \ref{fig:tpcc warehouse=thread} shows that all the systems exhibited scalability for a low contention case, and
 Fig. \ref{fig:tpcc warehouse=1} and \ref{fig:tpcc warehouse=4} show that they exhibited less efficiency for high contention cases.
 This is consistent with prior studies (Figs. 4 and 9 for a study on TicToc~\cite{yu2016tictoc}, Fig. 5 for a study on Cicada~\cite{lim2017cicada}, and Figs. 2 and 3 for a study on 1000 cores paper~\cite{10.1145/3399666.3399910}).
 Fig. \ref{fig:TPC-C-NP-NoInsert (Full Scale).} shows that DBx1000 outperformed CCBench in all cases. One of the reasons for this is the different indexes used.
 CCBench used Masstrees, and DBx1000 used hash tables.

 When the warehouse count was equal to the thread count, \seqsplit{DBx1000(C)} exhibited the best performance, as illustrated in Fig. \ref{fig:tpcc warehouse=thread}, when the thread count was less than or equal to 96.
 When the thread count was more than 96, it was not run owing to an ASSERT failure or segmentation fault. The increase in the {\tt NMaxCores} parameter did not solve this issue.
 Fig. \ref{fig:TPC-C-NP-NoInsert (Small Scale).} showed that all three platforms scaled in all settings when the thread count was no more than 20.

 %
 %
 {\bf NP-Insert: }
 For \seqsplit{DBx1000(C)}, we set two parameters to {\tt true},
 {\tt TPCC\_INSERT\_ROWS} and {\tt TPCC\_INSERT\_INDEX}.
 We uncommented the code of the insertion operations for DBx1000.
 We set the corresponding command-line argument for CCBench to omit insertion operations.
 The results are shown in Figs. \ref{fig:TPC-C-NP-Insert (Full Scale).} and \ref{fig:TPC-C-NP-Insert (Small Scale).}.
 The performance of DBx1000 for {\sf NP-Insert} was overwhelmingly worse than that of {\sf NP-NoInsert}.
 This should be due to the insertion operation.
 This degradation was reported by Bang et al.~\cite{10.1145/3399666.3399910} (Fig. 9), who improved the workload by weaving modern optimizations~\cite{kimura2015foedus,CompileVector,STOv2} into DBx1000.
 The codes were not publicly available.
 CCBench and DBx1000 exhibited similar trends in performance in all cases.
 CCBench outperformed DBx1000 under this setting.
 \seqsplit{DBx1000(C)} exhibited a mysterious performance when the warehouse count was set to 1 or 4 as shown in Figs. \ref{fig:tpcc warehouse=1, INSERT}, \ref{fig:tpcc warehouse=4, INSERT}, \ref{fig:tpcc warehouse=1(small), INSERT}, and \ref{fig:tpcc warehouse=4(small), INSERT}.
 We could not determine the reason for this behavior.
 %

 The performance of NP-Insert was worse than that of NP-NoInsert because of additional insert operations.
 CCBench exhibited approximately 14 Mtps for NP-NoInsert and approximately 8 Mtps for NP-Insert, with 224 threads and 224 warehouses
 in Figs.~\ref{fig:TPC-C-NP-NoInsert (Full Scale).}(c) and \ref{fig:TPC-C-NP-Insert (Full Scale).}(c), respectively.
 Insert operations require an index traversal as well as memory allocation for index nodes and record data, which typically cause page faults in the operating system.
 Therefore, they are considered to be expensive compared with other operations.

 {\bf Impact of Index to Performance: }
 %
 %
 The difference in indexes produces a difference in performance.
 The access cost of the tree indexes is higher than that of the hash indexes in theory.
 To understand the impact of the Masstree index on performance,
 we measured the latency breakdown of the transactions in the {\sf NP-Insert} workload of CCBench on 224 threads and 224 warehouses using the Linux perf tool.
 As shown in Fig. \ref{fig:breakdown}, execution times of 68.4\% and 58.9\%
 were spent for {\tt New-Order} and {\tt Payment} transactions on the search, update, and insert.
 These operations need to find a record, and require frequent Masstree traversals.
Because the size of the Masstree node was a few hundred bytes, its traversal decreased the spatial locality of the memory accesses; thus, the cache miss tended to increase.
{\sf NP} can be executed using hash indexes, and in such a case, the ratio of search, update, and insert to the execution time would be significantly reduced, and the performance of {\sf NP} would improve.

\subsubsection{Varying Warehouse Count}
We evaluated {\sf NP-NoInsert} and {\sf NP-Insert} varying the warehouse counts.
The settings are the same as those in Appendix \ref{sec:Varying Thread Count}.
The results for {\sf NP-NoInsert} and {\sf NP-Insert} were shown in
Figs.
\ref{fig:Varying warehouse count, NP-NoInsert}
and
\ref{fig:Varying warehouse count, NP-Insert}.
All results of \seqsplit{DBx1000(C)} were measured with 96 threads due to errors.
The results showed that all three systems tended to scale.
For {\sf NP-NoInsert}, DBx1000 outperformed both \seqsplit{DBx1000(C)} and CCBench as shown in Fig. \ref{fig:Varying warehouse count, NP-NoInsert}.
A result of a similar experiment is shown in Fig. 5 for the study on TicToc~\cite{yu2016tictoc},
and it exhibited scalability, which is consistent with this result.
For {\sf NP-Insert}, DBx1000 underperformed both \seqsplit{DBx1000(C)} and CCBench, as shown in Fig. \ref{fig:Varying warehouse count, NP-Insert}.

 \subsubsection{Difference between NP and Full-mix}
 \label{sec:Diff}
Transactions of TPC-C include the {\tt Order-Status}, {\tt Delivery}, and {\tt Stock-Level} in addition to {\tt New-Order} and {\tt Payment}.
We did not obtain the results in this study.
Under a full-mix workload, which includes all of these factors, its performance deteriorates to a fraction of the {\sf NP}.
Figs. 4 and 5 for the original study on Cicada~\cite{lim2017cicada} show the results of both full-mix and {\sf NP}, respectively.
The result in Fig. 4 was approximately $1/3$ of that in Fig. 5.
For TPC-C, the {\tt New-Order} and {\tt Payment} transactions account for 88\%, and the remaining three account for only 12\%.
The reason for the performance degradation was because the {\tt Delivery} transaction contained many record accesses compared to the {\tt New-Order} and {\tt Payment} transactions.
Owing to the existence of long transactions, many conflicts will occur for small warehouse cases, which would deteriorate the performance. For many warehouses, we expect the abort ratio would be low because Kimura~\cite{kimura2015foedus} determined that the ratio was 0.12\% for the warehouse count was equal to the thread count.
The effect of such long transactions was described in $\S$\ref{sec:limitation-of-current-approaches} for the version lifetime management.
The issues related to long transactions have rarely been discussed thus far, and remain a research problem.

\section{Evaluation of a TPC-C Full-mix}
\label{sec:Evaluation of a TPC-C Full-mix}
TPC-C full-mix consists of five transactions, namely {\tt New-Order}, {\tt Payment}, {\tt Delivery}, {\tt Stock-Level}, and {\tt Order-Status}.
Their percentages are 45, 43, 4, 4, and 4 \%, respectively.
We re-implemented Silo for the TPC-C full-mix in CCBench with our own implementation of Masstree.
%
%
%
Currently, the code for this version of CCBench is not publicly available at our repository~\cite{thawkccbench}.
However, we will release them soon.

\subsection{Design}
We prepared nine tables that were accessed using primary keys.
They include {\tt WAREHOUSE}, {\tt CUSTOMER}, {\tt NEW-ORDER}, {\tt DISTRICT},
{\tt ORDER}, {\tt ORDERLINE}, {\tt ITEM}, {\tt STOCK}, and {\tt HISTORY}.
The {\tt HISTORY} table uses surrogate keys, which are generated in a scalable manner.
We provided two bytes for worker thread ID and increased counters for the key to avoid contentions.

We prepared secondary tables for {\tt CUSTOMER} and {\tt ORDER}.
As for the {\tt CUSTOMER} secondary table,
the key consists of warehouse ID, district ID, and c\_last,
and the value consists of c\_first and the list of primary keys.
As for the {\tt ORDER} secondary table,
the key consists of warehouse ID, district ID, customer ID, and order ID, and the value is the primary key for the {\tt ORDER} table.

\subsection{Phantoms}
\label{sec:Phantoms}
In the TPC-C full-mix workload, the phantom problem can occur in the following cases.
\begin{itemize}
\item {\tt ORDER} secondary table.

  Scan operations in the {\tt Order-Status} transaction and insert operations in the {\tt New-Order} transaction can occur concurrently.

\item {\tt NEW-ORDER} table.

  Scan operations in the {\tt Delivery} transaction and insert operations in the {\tt New-Order} transaction can occur concurrently.
\end{itemize}

However, the phantom problem does not occur in the following cases.
\begin{itemize}

 \item {\tt ORDERLINE} table in the {\tt Stock-Level} transaction. 
   
   The {\tt New-Order} transaction inserts new records into the {\tt ORDERLINE} table, and the {\tt Stock-Level} transaction scans the {\tt ORDERLINE} table.
       %
   However, the range of keys scanned by the {\tt Stock-Level} transaction is different from the keys inserted by the {\tt New-Order} transaction.
   The {\tt New-Order} transaction increments {\sf d\_next\_id} and a record with the key is inserted into the {\tt ORDERLINE} table, while the {\tt Stock-Level} transaction scans keys with less than the {\sf d\_next\_id} from the {\tt ORDERLINE} table. 
       The key of the {\tt ORDERLINE} table consists of warehouse ID, district ID, order ID, and orderline ID.

 \item {\tt ORDERLINE} table in the {\tt Order-Status} transaction. 
       
       The {\tt New-Order} transaction inserts new records into the {\tt ORDERLINE} table, and the {\tt Order-Status} transaction scans the table. 
       However, it scans records inserted by past {\tt New-Order} transactions and not by concurrently running ones. 
       
 \item {\tt ORDERLINE} table in the {\tt Delivery} transaction. 
       
       The {\tt Delivery} transaction processes the oldest record that has not yet delivered. Therefore, the scan for the {\tt ORDERLINE} table is executed over past records.
       %
       %
       The {\tt New-Order} transaction inserts new records, and thus these scan and insert operations do not conflict.
       
 \item {\tt CUSTOMER} secondary table in the {\tt Payment} transaction. 

   No transactions insert into the {\tt CUSTOMER} table.
       
 \item {\tt CUSTOMER} secondary table in the {\tt Order-Status} transaction.

   No transactions insert into the {\tt CUSTOMER} table.
\end{itemize}

\subsection{Evaluation}

We implemented a TPC-C full-mix dealing with phantom problems, following the scheme described in Section \ref{sec:Phantoms}.
We evaluated the throughput and abort ratio varying thread count and the results are shown in Figures
 \ref{fig:Varying thread count, throughput, fullmix.} and 
 \ref{fig:Varying thread count, abort ratio, fullmix.}, respectively.
%
%
 %
 Each line in these figures represents the different settings of the warehouse count (1, 7, 28, 56, 84, 112, 140, 168, 196, 224, and the same number with the thread count).
 Figure \ref{fig:Varying thread count, throughput, fullmix.} shows that 
 the throughput linearly increases as the thread count increases, if the warehouse count is sufficient for the thread count.
The performance curve deteriorates when the thread count exceeds 112.
This is because our environment has only 112 physical cores among the 224 logical cores.
As illustrated in Figure
 \ref{fig:Varying thread count, abort ratio, fullmix.}, 
 the abort ratio increases if the warehouse count is insufficient for the thread count.
 This is because
 the contention increases with the increasing number of threads that access the same warehouse.

\begin{figure}[tb]
 \centering
 \includegraphics[scale=0.55]{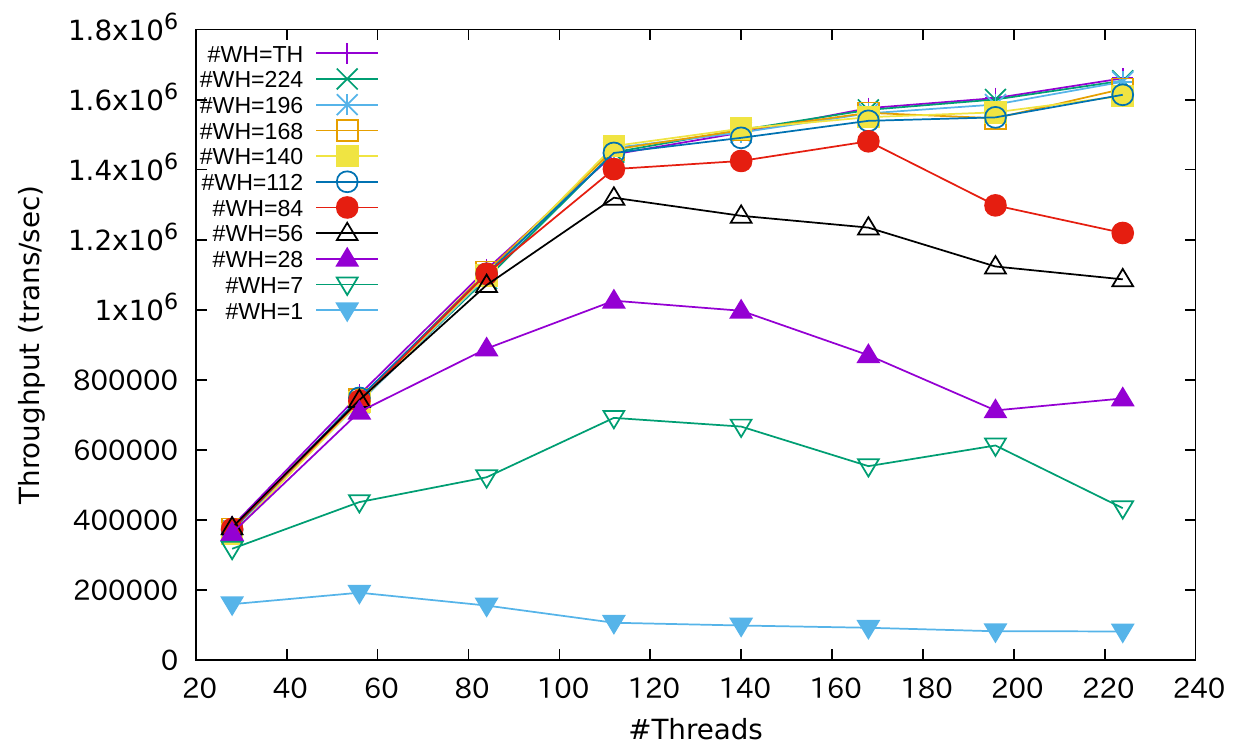}
 \caption{TPC-C full-mix: Throughput against thread count.}
 \label{fig:Varying thread count, throughput, fullmix.}
%
 \centering
 \includegraphics[scale=0.55]{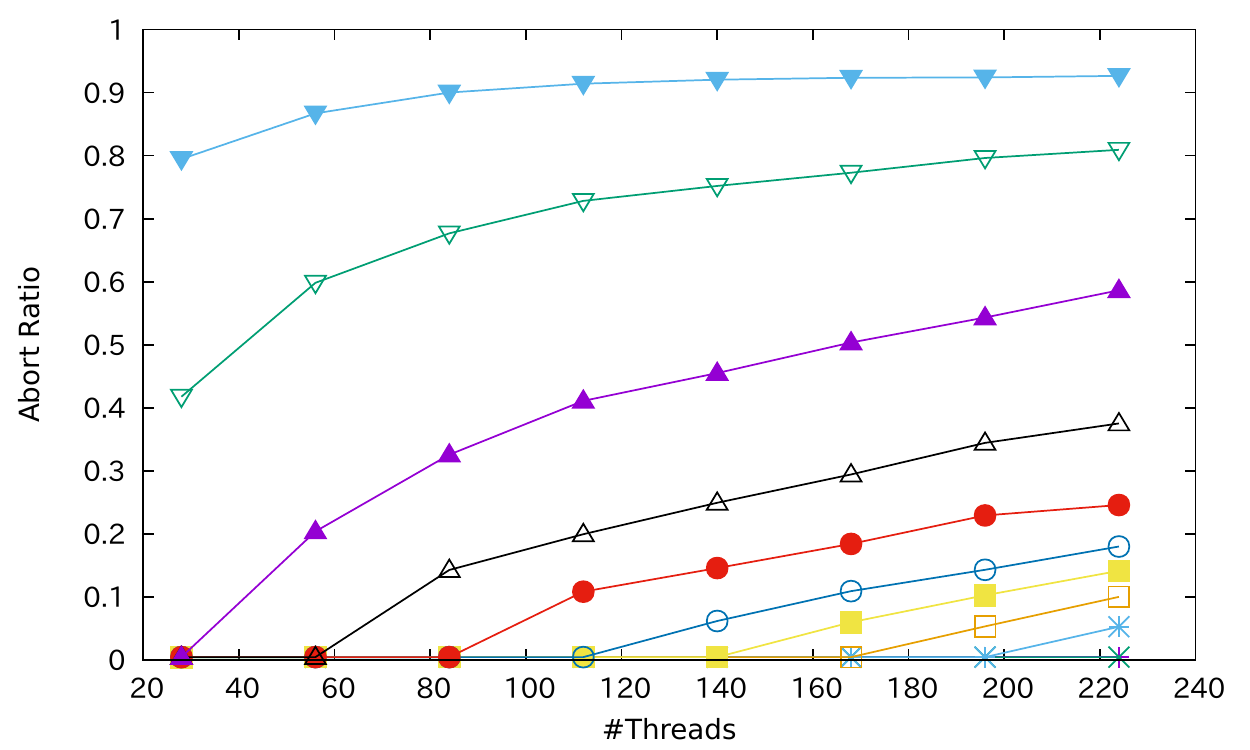}
 \caption{TPC-C full-mix: Abort ratio against thread count.}
 \label{fig:Varying thread count, abort ratio, fullmix.}
\end{figure}
%


As illustrated in Figure
\ref{fig:Varying warehouse count, fullmix with 224 threads.},
the throughput increases and the abort ratio decreases as the warehouse count increases when the number of worker threads is fixed at 224.
This is because of the increasing abort ratios caused by the contentions.
\begin{figure}[tb] 
 \centering
 \includegraphics[scale=0.55]{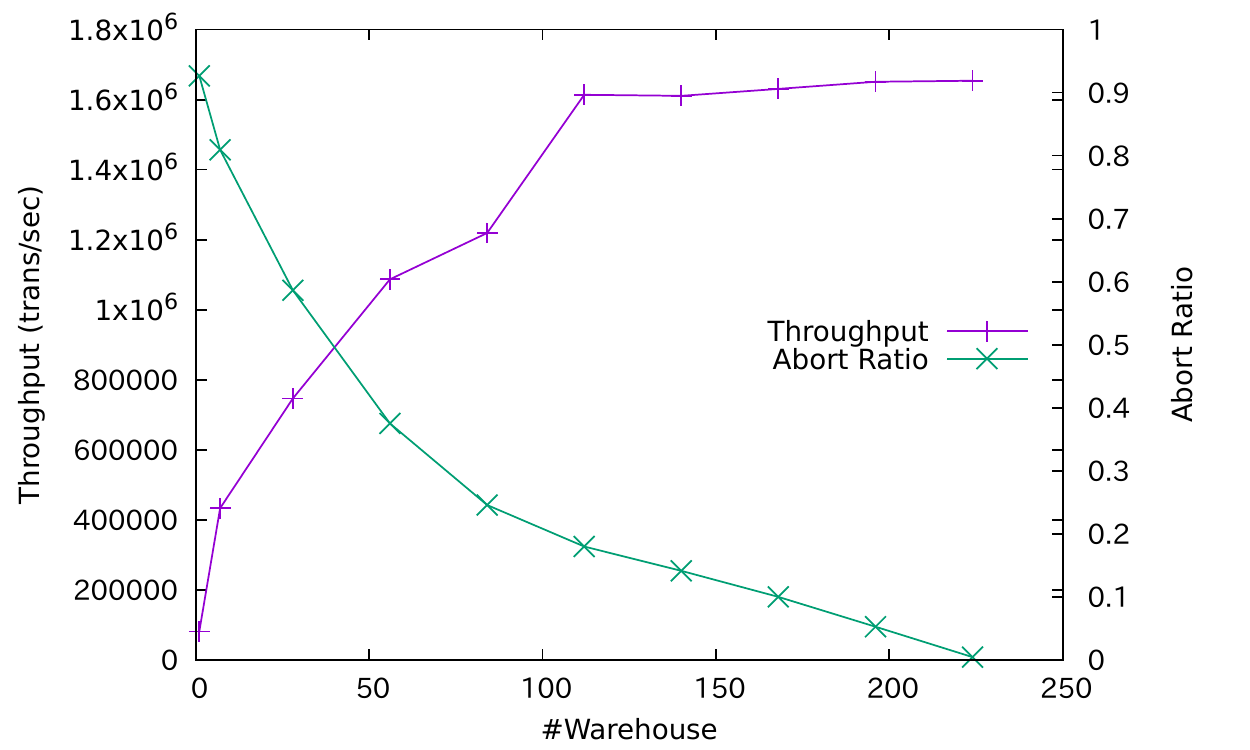}
 \caption{TPC-C full-mix: Throughput and abort ratio with 224 threads.}
 \label{fig:Varying warehouse count, fullmix with 224 threads.}
\end{figure}

 \subsection{Summary of TPC-C}
 TPC-C is an important and realistic benchmark that includes phantom occurrences by concurrent scan and insert operations.
 CCBench currently supports the TPC-C benchmark only 
 for the Silo protocol.
 This Appendix validates the correctness of its implementation by comparing CCBench with DBx1000~\cite{dbx1000} and \seqsplit{DBx1000(C)}~\cite{dbx1000_cicada_git} for TPC-C NP, and scalability for TPC-C full-mix.
 The results of the experiments demonstrated that the implementation of CCBench is appropriate.
 In addition, different analysis platforms exhibited different behaviors, even for the same workload, depending on the design and implementation of the systems. 
 We claim Insight 5 again; it is difficult to find deep knowledge using multiple reference implementations or platforms. We believe that CCBench will help in such an analysis and related research in the many-core architecture era.

 %
 %
 An evaluation of the TPC-C full-mix for protocols other than Silo remains as future work.
%
 %
 For TicToc, Cicada, MOCC, and 2PL-NoWait, the technique presented in this Appendix can be used.
 For 2PL-Wait, next-key locking with Masstree can be used.
 For SI, no extension is necessary for phantom avoidance because each transaction under SI reads a fixed snapshot, and the phantom problem does not occur.

\end{appendix}

\section*{Acknowledgments}
We thank Sho Nakazono for valuable advices on implementing TPC-C full-mix benchmark.
We thank Hideaki Kimura for valuable advices on transaction processing and encouragements.
We are grateful to the anonymous associate editor, the reviewers, and the shepherd for their insightful comments and valuable feedback.
We thank DBx1000~\cite{dbx1000} team,
Cicada team~\cite{dbx1000_cicada_git} and
tpcc-runner team~\cite{tpcc_runner_git}
for publishing code.
This work was supported by JST JPMJCR1414, JSPS 19H04117, and the New Energy and Industrial Technology Development Organization (NEDO).

\fi
\end{document}